\documentclass[12pt]{article}
\pdfoutput=1
\usepackage{amsmath,amsfonts,amssymb,graphicx,color,bbm,tikz,float}
\usepackage{comment}
\usepackage[nosort]{cite}
\usepackage{subfigure}
\usetikzlibrary{calc,positioning}
\usetikzlibrary{patterns,arrows,decorations.pathreplacing}
\usepackage{caption}
\usepackage{ulem}
\tikzset{>=stealth}
\usepackage{lipsum}
\usepackage{hyperref}

\newcommand\blfootnote[1]{%
  \begingroup
  \renewcommand\thefootnote{}\footnote{#1}%
  \addtocounter{footnote}{-1}%
  \endgroup}

\makeatletter\@addtoreset{equation}{section}\makeatother
\newcommand{\be}{\begin{equation}}
\newcommand{\ee}{\end{equation}}
\def\beq{\begin{equation}}
\def\eeq{\end{equation}}
\newcommand{\bea}{\begin{eqnarray}}
\newcommand{\eea}{\end{eqnarray}}

\newcommand{\bra}[1]{{\left< {#1} \right|}}
\newcommand{\ket}[1]{{\left| {#1} \right>}}

\renewcommand{\title}[1]{\vbox{\center\LARGE{#1}}\vspace{3mm}}
\renewcommand{\author}[1]{\vbox{\center#1}\vspace{3mm}}

\newcommand{\email}[1]{\vbox{\center\tt#1}\vspace{3mm}}

\begin{document}
\begin{titlepage}
\begin{center}

{\large {\bf Braiding quantum gates from partition algebras} }

\author{Pramod Padmanabhan,$^a$ Fumihiko Sugino,$^b$ Diego Trancanelli$^{c,\star}$}\blfootnote{${}^\star$ On leave of absence from the Institute of Physics at the University of S\~ao Paulo, S\~ao Paulo, Brazil.}

\vskip -1cm
{$^a${\it Center for Theoretical Physics of Complex Systems,\\ Institute for Basic Science, Daejeon, South Korea}}
\vskip0.1cm
{ $^b${\it Center for Theoretical Physics of the Universe,\\
Institute for Basic Science, Daejeon, South Korea} 
\vskip0.1cm
$^c${\it Dipartimento di Scienze Fisiche, Informatiche e Matematiche, \\
Universit\`a di Modena e Reggio Emilia, via Campi 213/A, 41125 Modena, Italy \\ \& \\
INFN Sezione di Bologna, via Irnerio 46, 40126 Bologna, Italy}}
\email{pramod23phys, fusugino, dtrancan@gmail.com}

\vskip 1cm 
\end{center}

\abstract{
\noindent 
Unitary braiding operators can be used as robust entangling quantum gates. We introduce a solution-generating technique to solve the $(d,m,l)$-generalized Yang-Baxter equation, for $\frac{m}{2}\leq l \leq m$, which allows to systematically construct such braiding operators. This is achieved by using 
partition algebras, a generalization of the Temperley-Lieb algebra encountered in statistical mechanics. We obtain families of unitary and non-unitary braiding operators that generate the full braid group. Explicit examples are given for a 2-, 3-, and 4-qubit system, including the classification of the entangled states generated by these operators based on Stochastic Local Operations and Classical Communication.
}

\end{titlepage}
\tableofcontents 

\section{Introduction}

The fragile nature of quantum entanglement is a central issue in quantum computing, which can in principle be alleviated by the use of topology. Drawing inspiration from the Aravind hypothesis \cite{pk, sug, quinta}, it has been proposed that braiding operators -- operators that obey braiding relations and create knots from unlinked strands -- could be thought of as quantum entanglers, {\it i.e.} gates that create entanglement from product states \cite{lh1, lh2, lh3, lh4, lh5}. These initial studies about the relation between entangling gates and knots were then pushed forward in \cite{ste, r1, r2, wk, w2}, paving the way to the proposal of topological quantum circuits with gates given by braiding operators \cite{lh6, lh7}. It is expected that a physical realization of these braiding operators/entangling gates could be obtained using anyons.

One way to get braiding operators is by solving the {\it parameter-independent} Yang-Baxter equation (YBE), which we briefly review.\footnote{For details about the case in which the YBE depends on a so-called spectral parameter, see {\it e.g.} \cite{pdd} and references therein.} The YBE is an operator equation for an invertible matrix, $R : V\otimes V\rightarrow V\otimes V$, given by
\begin{equation}\label{ybe}
\left(R\otimes I\right)\left(I\otimes R\right)\left(R\otimes I\right) = \left(I\otimes R\right)\left(R\otimes I\right)\left(I\otimes R\right),
\end{equation}
where $V$ is a $d$-dimensional complex vector space and $I$ is the identity operator on $V$. 
We use the terms {\it Yang-Baxter operator} and {\it R-matrix} interchangeably for the operator $R$. 
Solutions to (\ref{ybe}) for some cases are presented in \cite{hietarinta2,hietarinta,dye}.

The $R$-matrix can be seen as a generalization of the permutation operator that swaps two vector spaces. This point of view is useful if one notices that the $R$-matrices can be used to construct representations of the {\it Artin braid group} $B_n$ on $n$-strands, with generators $\sigma_i$ satisfying
\begin{eqnarray}
\sigma_i\sigma_{i+1}\sigma_i & = & \sigma_{i+1}\sigma_i\sigma_{i+1}, \label{brelation}\\
\sigma_i\sigma_j & = & \sigma_j\sigma_i,~~|i-j|>1,\label{fc}
\end{eqnarray}
for $i=1,\ldots, n-1$. The first relation above is called the {\it braid relation}, whereas the second relation is the {\it far-commutativity} condition. Representations for $\sigma_i$ can be constructed out of the $R$-matrices that solve \eqref{ybe} as follows
\begin{equation}
\rho(\sigma_i) = I^{\otimes i-1}\otimes R_{i, i+1}\otimes I^{\otimes n-i-1}.
\end{equation}
Notice that this representation satisfies far-commutativity trivially. This implies that every $R$-matrix that solves \eqref{ybe} can be used to construct a representation of the braid group, denoted a {\it braiding operator}. 

The distinction between $R$-matrices and braiding operators become essential when introducing a natural generalization of the YBE \cite{r1,r2} which involves two extra parameters, $m$ and $l$. The linear invertible operator $R : V^{\otimes m} \rightarrow V^{\otimes m}$ now acts on $m$ copies of the $d$-dimensional vector space $V$ with $l$ identity operators, and obeys
\begin{equation}\label{gybe}
\left(R\otimes I^{\otimes l}\right)\left(I^{\otimes l}\otimes R\right)\left(R\otimes I^{\otimes l}\right) = \left(I^{\otimes l}\otimes R\right)\left(R\otimes I^{\otimes l}\right)\left(I^{\otimes l}\otimes R\right),
\end{equation}
prompting the notation $(d,m,l)$-gYBE, as used in \cite{rchen}. We dub this generalized $R$-operator as either the {\it generalized Yang-Baxter operator} or the {\it generalized $R$-matrix}. This generalization is important for quantum information processes that involve more than two qubits. 

Unlike the $R$-matrix that solves the YBE in \eqref{ybe}, not all generalized $R$-matrices that solve the $(d,m,l)$-gYBE in \eqref{gybe} provide a representation of the braid group, as they do not always satisfy the far-commutativity condition in \eqref{fc}. However, for the cases when $2l \geq m$ (assuming $l< m$) far-commutativity is trivially satisfied, just as in the case of the Yang-Baxter operators. This is seen through the representations of the braid group given in terms of the generalized $R$-matrices by
\begin{equation}
\rho(\sigma_i) = \left(I^{\otimes l}\right)^{\otimes i-1}\otimes R_{i,\cdots, i+m-1}\otimes \left(I^{\otimes l}\right)^{\otimes n-i-m+1}. 
\end{equation}
We will then be interested in finding the generalized Yang-Baxter operators that satisfies the $(d,m,l)$-gYBE when $2l \geq m$, thus automatically leading to representations of the braid group.\footnote{A different approach to representations to the braid group is discussed in \cite{GKRZ}.} 

The $(d,m,l)$-gYBE in \eqref{gybe} involves $d^{2m+2l}$ cubic polynomial relations for $d^{2m}$ unknowns (the entries of the generalized $R$-matrix) and is in general hard to solve. In this work we use algebraic methods to solve for the $R$-matrices and generalized $R$-matrices using {\it partition algebras} \cite{Jo,Ma1,pm,Ma3,Ma4,MR}. We obtain families of both unitary and non-unitary operators, with the former finding use as quantum gates in quantum computing and the latter being useful to investigate topological aspects of the gYBE that include the study of knot invariants. We focus on the quantum entangling aspects of these operators.

The paper is structured as follows. Set partitions and partition algebras are reviewed in Sec. \ref{pa}, along with representations (the {\it Qubit} and the {\it Temperley-Lieb representations}) of their modified versions. In Sec. \ref{Req} we recall the equivalence classes of the Yang-Baxter operators and discuss how they relate to the notion of {\it Stochastic Local Operations and Classical Communication} (SLOCC) classes of entangled quantum states in quantum information theory. Our main results are in Sec. \ref{Rmatrices}, where we obtain and discuss in detail $R$-matrices for the 2-, 3-, and 4-qubit cases. We also study the SLOCC classes of entangled states generated by these matrices. The structure of these generalized $R$-matrices allows to find an algorithm to systematically generate solutions of the $(d,m,l)$-gYBE. There are three kinds of generalized $R$-matrices known in the 3-qubit case \cite{r1, rchen, wk}. In Sec.~\ref{compR} we show that the 3-qubit solutions obtained in this paper are inequivalent to the known solutions. We conclude with some open questions and an outlook in Sec. \ref{out}. 

\section{Set partitions and partition algebras} \label{pa}

We review the notion of set partitions and partition algebras following \cite{par}. We present just the bare minimum needed in this work, pointing the reader to that reference for more details.  The elements of set partition, denoted $A_k$, are the partitions of two copies of a set: $\{1,2,\cdots, k\}$ and $\{1', 2', \cdots, k'\}$. As an example consider the following diagram showing the partition of a set with $k=7$, this represents the set partition $\left\{\left\{ 1,3,5,4',5'\right\}, \left\{2,3'\right\},\left\{4,6,7,6'\right\},\left\{1',2'\right\}, \left\{7'\right\}\right\}.$ 
\begin{figure}[h!]
	\begin{center}
		\includegraphics[scale=1]{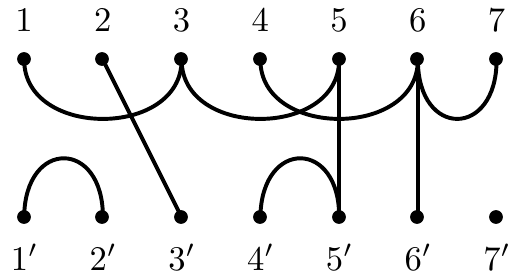}
	\caption{A diagram representing $ \left\{\left\{ 1,3,5,4',5'\right\}, \left\{2,3'\right\},\left\{4,6,7,6'\right\},\left\{1',2'\right\}, \left\{7'\right\}\right\}\in A_7$.}
	\label{pelem}
	\end{center}
\end{figure}
In the diagram, vertices $i$ and $j$ are connected by a path if $i$ and $j$ belong to the same block of the partition. 
Note that the diagram for a given element of $A_k$ is not unique. 
For example the diagram in Fig. \ref{pelem2} also represents the same element represented in the diagram Fig. \ref{pelem}.
\begin{figure}[h!]
	\begin{center}
		\includegraphics[scale=1]{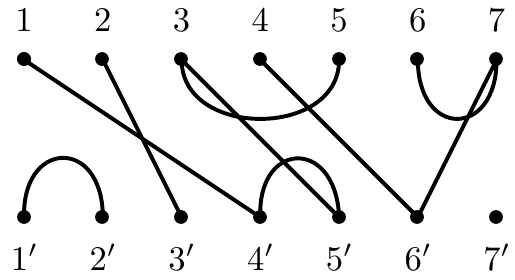}
	\caption{Another diagram representing the same element of $A_7$ shown in Fig. \ref{pelem}.}
	\label{pelem2}
	\end{center}
\end{figure}

\begin{figure}[h!]
	\begin{center}
		\includegraphics[scale=1]{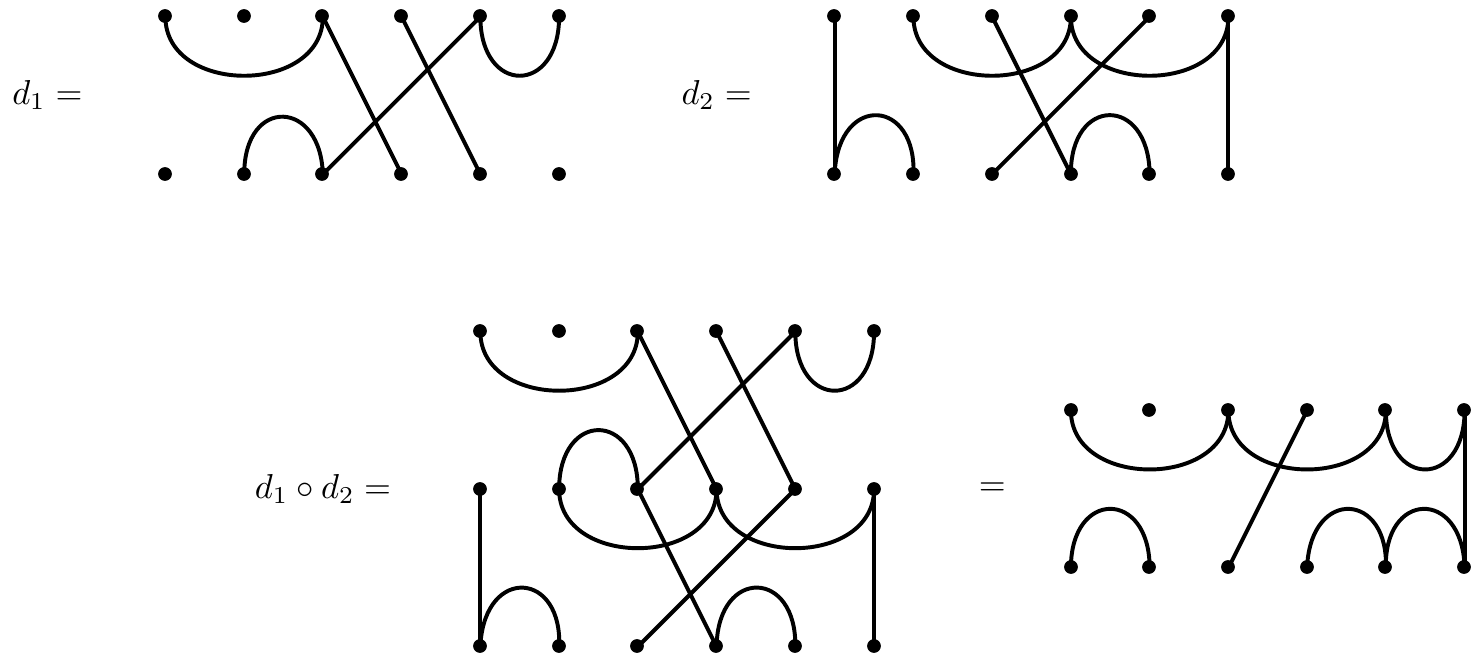}
	\caption{Composition of elements of $A_6$.}
	\label{pcomp6}
	\end{center}
\end{figure}

To compose two diagrams, $d_1\circ d_2$, place $d_1$ above $d_2$ and then trace the lines to obtain the new partition. An example of such a composition is given for the case of $k=6$ in Fig. \ref{pcomp6}.

\begin{figure}[h!]
	\begin{center}
		\includegraphics[scale=1]{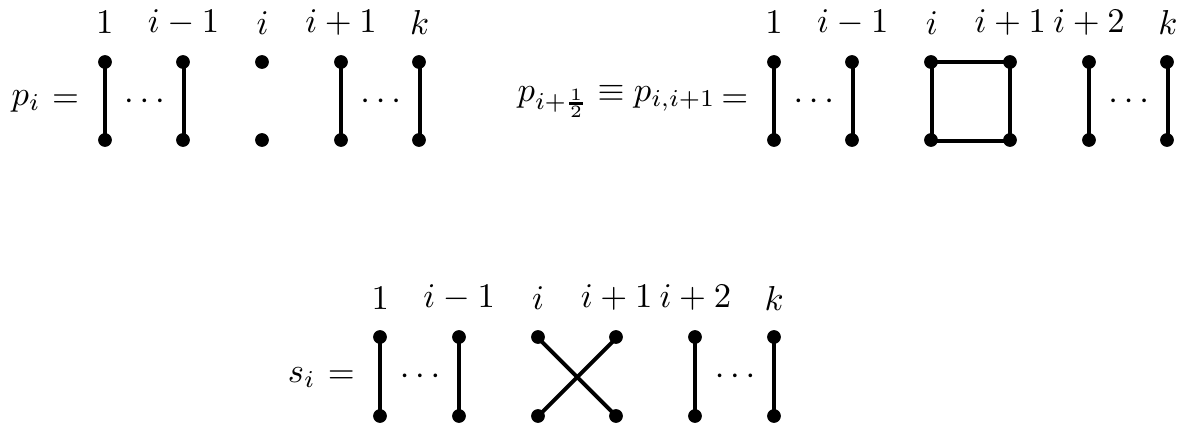}
	\caption{Generators of $A_k$.}
	\label{pgen}
	\end{center}
\end{figure}

\begin{figure}[h!]
	\begin{center}
		\includegraphics[scale=1]{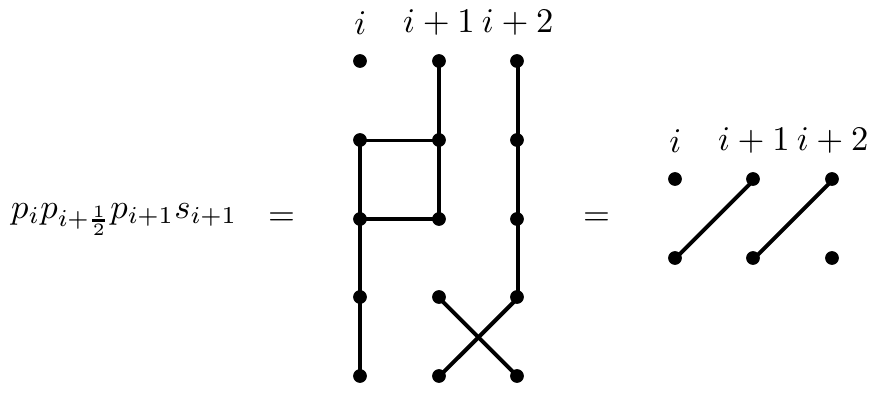}
	\caption{Example of composition in $A_k$. The nodes $1,\cdots i-1, i+3,\cdots, k$ on which the generators act trivially are suppressed.}
	\label{pcomp}
	\end{center}
\end{figure}

The elements of $A_k$ are generated by successive compositions of 
\begin{eqnarray}\label{p}
&& p_i, \;\;\;\;  \qquad\textrm{for}~i\in\{1, \cdots, k\},
\\
\label{p12}
&& p_{i+\frac{1}{2}},\qquad \textrm{for}~i\in\{1,\cdots, k-1\},
\\
\label{s}
&& s_i,  \;\;\;\;  \qquad \textrm{for}~i\in\{1,\cdots, k-1\},
\end{eqnarray}
whose action can be represented diagrammatically, see Fig. \ref{pgen}.
An example of composition of these generators is shown in Fig. \ref{pcomp}. 
Using these diagrams one can easily verify that the generators satisfy the following relations
\begin{equation}\label{squaring}
p_i^2=p_i,\qquad p^2_{i+\frac{1}{2}}=p_{i+\frac{1}{2}},
\end{equation}
\begin{equation}\label{prelations1}
p_ip_{i\pm\frac{1}{2}}p_i=p_i,\qquad p_{i\pm\frac{1}{2}}p_ip_{i\pm\frac{1}{2}}=p_{i\pm\frac{1}{2}},
\end{equation}
\begin{equation}\label{prelations2}
p_ip_j=p_jp_i,~\textrm{for}~|i-j|>\frac{1}{2},
\end{equation}
\begin{equation}\label{permRelations}
s_i^2=1,\qquad s_is_{i+1}s_i = s_{i+1}s_is_{i+1},\qquad s_is_j=s_js_i,~\textrm{for}~|i-j|>1.
\end{equation}
Here and below, we simply write $d_1 \circ d_2$ as $d_1d_2$ for notational simplicity.

\begin{figure}[h!]
	\begin{center}
		\includegraphics[scale=1]{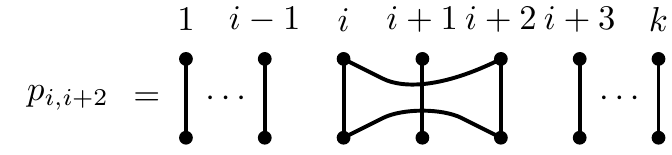}
	\caption{ The element $p_{i, i+2}$.}
	\label{pi2}
	\end{center}
\end{figure}

Note that $p_i$ and $p_{i+\frac{1}{2}}$ generate planar diagrams. Non-planarity is introduced by the permutation group generators $s_i$. The mixed relations are 
\begin{equation}\label{sprelations1}
s_ip_ip_{i+1}=p_ip_{i+1}s_i=p_ip_{i+1},\quad s_ip_is_i=p_{i+1},\quad s_ip_{i+j}=p_{i+j}s_i,~\textrm{for}~j\neq 0, 1,
\end{equation}
and 
\begin{equation}\label{sprelations2}
s_ip_{i+\frac{1}{2}}=p_{i+\frac{1}{2}}s_i=p_{i+\frac{1}{2}},\quad s_is_{i+1}p_{i+\frac{1}{2}}s_{i+1}s_i=p_{i+\frac{3}{2}},\quad s_ip_{i+j}=p_{i+j}s_i,~\textrm{for}~j\neq -\frac{1}{2}, \frac{3}{2}.
\end{equation}
To emphasize that $s_i$ swaps the elements on the vector spaces at sites $i$ and $i+1$, one could also write it as $s_{i,i+1}$, but we will stick to the notation $s_i$ to avoid cluttering. 
The second relations in \eqref{sprelations1}-\eqref{sprelations2} can be understood as the fundamental property of the permutation operator:
\begin{eqnarray}
s_ip_is_i  =  p_{i+1},\qquad  \label{sprelations3}
s_{i+1}p_{i+\frac{1}{2}}s_{i+1}  =  s_ip_{i+\frac{3}{2}}s_i, \label{sprelations4}
\end{eqnarray}
The index swapping is obvious in the first relation and it becomes obvious also in the second one, when one notices that $p_{i+\frac{1}{2}}$ has non-trivial support on sites $i$ and $i+1$ whereas $p_{i+\frac{3}{2}}$ has non-trivial support on sites $i+1$ and $i+2$. To make this more transparent, one can change notation by identifying $p_{i+\frac{1}{2}}$ with $p_{i, i+1}$ and $p_{i+\frac{3}{2}}$ with $p_{i+1, i+2}$,  prompting the definition 
\begin{equation}
s_{i+1}p_{i,i+1}s_{i+1}  =  s_ip_{i+1,i+2}s_i \equiv p_{i,i+2},
\end{equation}
whose diagrammatic representation is shown in Fig. \ref{pi2} and can be worked out using the composition laws in Fig. \ref{pcomp}.  
The figure suggests one can generalize the $p_{i+\frac{1}{2}}\equiv p_{i,i+1}$ operators to the cases with support on two arbitrary sites, $i$ and $i+j$, as
\begin{equation}\label{pij1}
p_{i,i+j} = 
s_{i+j-1}s_{i+j-2}\cdots s_{i+1} p_{i, i+1} s_{i+1}\cdots s_{i+j-2}s_{i+j-1},
\end{equation}
satisfying the relations 
\begin{eqnarray}
p_{i, i+j}^2 & = & p_{i,i+j}, \label{pijrelations1} \\
p_{i, i+j_1}p_{i, i+j_2} & = & p_{i, i+j_1}p_{i+j_1, i+j_2},~~j_1<j_2, \label{pijrelations2} \\
p_{i+l, i+j}p_{i, i+j} & = & p_{i, i+j}p_{i+l, i+j}  =  p_{i, i+l}p_{i+l, i+j} ,~~l<j, \label{pijrelations3}
\end{eqnarray}
which can be verified diagrammatically. Henceforth, we shall use $p_{i,i+1}$ instead of $p_{i+\frac{1}{2}}$.

Linear combinations of elements of $A_k$ with coefficients being complex numbers form the partition algebra $\mathbb{C}A_k(1)$. 

\subsection{Representations}
We use a slightly modified form of the relations in (\ref{squaring}) where we either scale one or both of the relations by a factor, $d$. We denote them asymmetric or symmetric scaling respectively. 
To this end we employ hermitian representations for the generators, which come in two kinds: the generators of the planar diagrams can be rescaled either asymmetrically {\it (Qubit representation)} or symmetrically {\it (Temperley-Lieb representation)}. 
Strictly speaking, these representations do not give the representations of the relations (\ref{squaring})-(\ref{sprelations2}), but the representations of their deformed versions. 

\paragraph{Qubit representation}  In this representation one of the relations satisfied by $p_{i,i+1}$ is modified to
\begin{equation}
p_{i,i+1}^2 = d~p_{i,i+1},
\label{norm_pi1/2}
\end{equation}
with the other relations in \eqref{squaring}, \eqref{prelations1} and \eqref{prelations2} unchanged. Here $d$ is the dimension of the local Hilbert space on the site $i$ acted upon by the generators of $A_k$. 

The relations in the non-planar part of $A_k$ involving $p_{i,i+1}$ and $s_i$, see \eqref{sprelations2}, and the relations in \eqref{pijrelations2} and \eqref{pijrelations3} are unchanged. The relation in \eqref{pijrelations1} is modified to 
\begin{equation}
p_{i, i+j}^2  = d~ p_{i,i+j},
\end{equation}
corresponding to the scaling of $p_{i,i+1}$. 

In this paper we deal with qubits and hence the case of $d=2$. The qudit realizations can be similarly obtained through an appropriate generalization. Qubit representations are given by
\begin{equation}\label{zrep1}
p_i =  \frac{1+Z_i}{2}, \qquad 
p_{i,j} = 1 + X_iX_j, \qquad 
s_i =  \frac{1 + X_iX_{i+1} + Y_iY_{i+1} + Z_iZ_{i+1}}{2},
\end{equation}
where 1 is the identity operator acting on the relevant Hilbert space and $X_i, Y_i, Z_i$ are the usual Pauli matrices acting on the qubit space on site $i$,
\begin{equation}
X = \left(\begin{array}{cc}0 & 1 \\ 1 & 0 \end{array}\right),\quad  Y = \left(\begin{array}{cc}0 & -\mathrm{i} \\ \mathrm{i} & 0 \end{array}\right),\quad Z = \left(\begin{array}{cc}1 & 0 \\ 0 & -1 \end{array}\right),
\end{equation}
written in the basis $\{\ket{0}$, $\ket{1}\}$ where $Z$ is diagonal.
Another representation which is unitarily equivalent to the above is given by
\begin{equation}\label{xrep1}
p_i =  \frac{1+X_i}{2}, \qquad 
p_{i,j} = 1 + Z_iZ_j, \qquad 
s_i =  \frac{1 + X_iX_{i+1} + Y_iY_{i+1} + Z_iZ_{i+1}}{2}. 
\end{equation}

The qubit representation gives the representation of the relations (\ref{squaring})-(\ref{sprelations2}) with the normalization of $p_{i,i+1}\equiv p_{i+\frac12}$ changed to 
(\ref{norm_pi1/2}) with $d=2$.

\begin{figure}[h!]
	\begin{center}
		\includegraphics[scale=1]{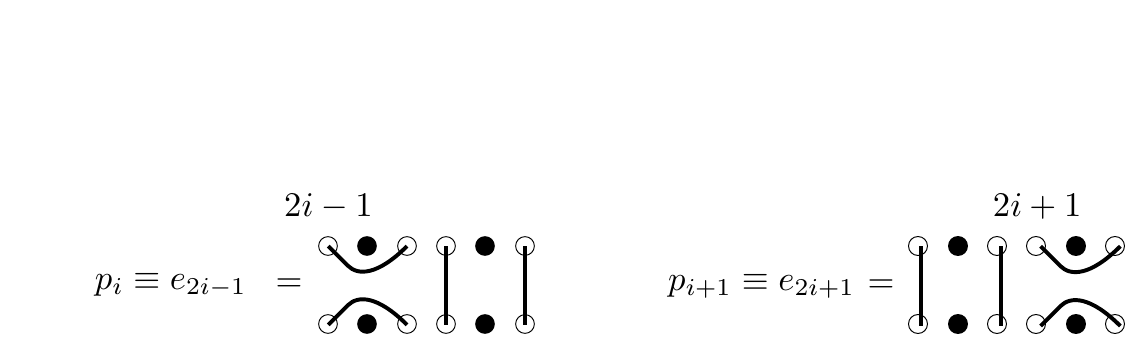}
	\caption{Temperley-Lieb representation of $p_i$ and $p_{i+1}$. 
	The white dots, obtained by doubling the original sites for $A_k$ (the black dots), are sites on which the Temperley-Lieb generators ($e_{2i-1}, e_{2i+1}$) act. }
	\label{tp}
	\end{center}
\end{figure}

\paragraph{Temperley-Lieb representation} Now both generators of the planar diagrams are rescaled by the same factor,
\begin{equation}
p_{i,i+1}^2 = (Q+Q^{-1})p_{i,i+1},\qquad p_i^2=(Q+Q^{-1})p_i, \qquad Q\in \mathbb{R}-\{0\}
\label{squaringTL}
\end{equation}
with the rest of the relations of the planar part of the partition algebra unchanged. The planar part of $A_k$ can be realized using the generators of the Temperley-Lieb algebra 
with doubled dimensions, $e_1,\cdots, e_{2k-1}\in {\rm TL}_{2k}$,
\begin{equation}\label{part-TL}
p_i = e_{2i-1}, \qquad p_{i,i+1}=e_{2i},
\end{equation}
which satisfy the relations in \eqref{prelations1}-\eqref{prelations2}, see \cite{par}. Notice the doubling of the number of sites in this realization, as shown in Fig. \ref{tp}.

\begin{figure}[h!]
	\begin{center}
		\includegraphics[scale=1]{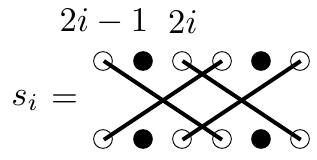}
	\caption{Temperley-Lieb representation of $s_i$ in (\ref{chichi}).}
	\label{ts}
	\end{center}
\end{figure}

In this representation the introduction of non-planarity through the permutation generators $s_i$ affects some of the mixed relations (\ref{sprelations1}) and (\ref{sprelations2}). Let us consider the case that $s_i$ is realized as an appropriate permutation operator given by 
\begin{equation}\label{chichi}
s_i = s_{2i-1, 2i+1}~s_{2i, 2i+2},
\end{equation}
or by the unitarily equivalent 
\begin{equation}
s_i = s_{2i-1, 2i+2}~s_{2i, 2i+1},
\end{equation}
with $s_{i,j}$ being the operator that swaps the indices $i$ and $j$. This realization lives on the doubled lattice as shown in Fig. \ref{ts}.
Using this diagram the partition algebra in \eqref{sprelations1} can be easily verified, whereas (\ref{sprelations2}) does not hold. 
Thus the Temperley-Lieb representation gives the representation of the relations (\ref{squaringTL}) and (\ref{prelations1})-(\ref{sprelations1}).

The doubling of the sites in this representation implies that one shall obtain $R$-matrices and generalized $R$-matrices on twice the number of sites, {\it i.e.} if one obtains the generalized $R$-matrix that solves the $(d,m,l)$-gYBE, then this representation will yield another solution that solves the $(d,2m,2l)$-gYBE. 

In section \ref{Rmatrices}, generalized $R$-matrices are constructed as linear combinations of the above representations of deformed set partitions that are analogous to elements of the partition algebras. 

\section{Equivalence classes of $R$-matrices and SLOCC classes} \label{Req}

In quantum information theory the idea of SLOCC was introduced to classify entangled states. It states that two quantum states are equivalent when there exists an {\it Invertible Local Operator} (ILO) that maps one state into the other:
\begin{equation}
\ket{\psi_1} = \left(A_1\otimes\cdots\otimes A_n\right)\ket{\psi_2},
\end{equation}
where the states $\ket{\psi_1}$ and $\ket{\psi_2}$ live in the Hilbert space $\otimes_{i=1}^n~\mathcal{H}_i$~\cite{dur}. 
$A_i$ is an ILO acting only at the site $i$. This equivalence relation appeals to the intuition of entangled states, as one expects local operations to not disturb the non-local entanglement property of the state.

One can also define an equivalence class of the $R$-matrices satisfying the parameter-independent YBE. 
To identify this class, one observes that if $R$ is a solution to the $(d,m,l)$-gYBE, then so are $\alpha R$ (with $\alpha$ a constant), $R^{-1}$, and $\left(A_1\otimes\cdots\otimes A_m\right)R\left(A_1^{-1}\otimes\cdots\otimes A^{-1}_m\right)$, where $A_1,\cdots, A_m$ is an ILO that also appears in the definition of the SLOCC classes of quantum states. We can now prove the following theorem:

\paragraph{Theorem} {\it Two entangling $R$-matrices $R_1$ and $R_2$, which are equivalent under ILO, produce entangled states of the same SLOCC class.}

\paragraph{Proof} $R_1$ produces an entangled state $\ket{E}$ acting on the product state $\ket{P}$,
\begin{equation}
R_1~\ket{P} = \ket{E}.
\end{equation}
By assumption, one can express $R_1 = AR_2A^{-1}$, where $A$ is an ILO, so that
\begin{equation}
AR_2A^{-1}~\ket{P} =  \ket{E}. 
\end{equation}
This means 
$
R_2A^{-1}~\ket{P} = A^{-1}~\ket{E}$, 
by definition, both $A^{-1}~\ket{P}$ and $A^{-1}~\ket{E}$ are in the same SLOCC classes as $\ket{P}$ and $\ket{E}$, respectively, hence proving the assertion.   ~~~~~~~~~~~~~~~~~~~~~~~~~~~~~~$\blacksquare$

This theorem naturally implies that if two $R$-matrices produce states of two different SLOCC classes, then they cannot be related by an ILO. However, the converse of the theorem is not always true: if two entangled states $\ket{E_1}$ and $\ket{E_2}$ belonging to the same SLOCC class are generated by two entangling $R$-matrices $R_1$ and $R_2$, respectively, then they need not be related by an ILO. One has in fact
\begin{eqnarray}
R_1~\ket{P_1} =  \ket{E_1}, \qquad
R_2~\ket{P_2}  =  \ket{E_2}. 
\end{eqnarray}
As $\ket{E_2}=A\ket{E_1}$ and $\ket{P_2}=L\ket{P_1}$, where $A$ and $L$ are two ILOs, one obtains
\begin{equation}\nonumber
A^{-1}R_2L~\ket{P_1} = \ket{E_1}.
\end{equation} 
Note that the ILOs $A$ and $L$ need not be the same.
For unitary $R$-matrices, this relation holds on all the product states that span the Hilbert space, so that one can identify
\begin{equation}\nonumber
R_1=A^{-1}R_2L.
\end{equation}
We shall use the definitions and this result to determine the classes of $R$-matrices.

\paragraph{2-qubit SLOCC classes}
There are two SLOCC classes in the 2-qubit case, the Bell state class and the product state class. 

\paragraph{3-qubit SLOCC classes}
There are six SLOCC classes in the 3-qubit case \cite{dur}. Two tri-partite entangled classes, GHZ state class and W state class, also denoted as $ABC$ to symbolize the three parties of the state. Three partially entangled state classes, $A-BC, AC-B, AB-C$ and finally the product state class, $A-B-C$. 

\paragraph{4-qubit SLOCC classes}
In the 4-qubit case, it was discussed in \cite{dur} that there are infinitely many SLOCC classes. 
Later, it was shown in \cite{VDMV} that there are nine families in the sense of nine different ways of entangling 4-qubits. 
On the other hand, it was reported in \cite{LLSS1,LLSS2} that the number of the families is eight instead of nine 
for genuinely entangled states.\footnote{The classification in \cite{VDMV} contains no genuinely entangled state with canonical state $\ket{0000}+\ket{0111}$. Due to this difference, 
\cite{LLSS1,LLSS2} does not contradict \cite{VDMV}.}  
Furthermore, it was discovered in \cite{LLHL} that the nine families in \cite{VDMV} further split into 49 SLOCC entanglement classes by looking at SLOCC invariant quantities.

\section{Generalized $R$-matrices}\label{Rmatrices}

The generators of the permutation group $s_i$ solve the $(d,2,1)$-gYBE. In fact, the transposition operators $s_{i,i+l}$ solve the $(d,m,l)$-gYBE, assuming $l\leq m$, with non-trivial support on the sites $i$ and $i+l$. The ansatze with non-trivial support on all the $m$ sites used in this paper are modifications of these transposition operators, with generators given by the planar part of the partition algebra. In the language of quantum gates, these ansatze are generalized SWAP gates. 

In the following we discuss the 2-qubit and 3-qubit cases in detail before writing down the answers for the 4-qubit case and outlining an algorithm for an arbitrary multi-qubit generalized $R$-matrix.  

\subsection{2-qubits}\label{2quR}

On two sites $i$ and $i+1$, there are various choices of the generators $p_i$, $p_{i+1}$, $p_{i,i+1}$ and $s_i$ to construct the $R$-matrices. We consider the different possibilities separately.

\paragraph{Using $s_i$, $p_i$ and $p_{i+1}$}

Consider the following ansatz for the Yang-Baxter operator with support on sites $i$ and $i+1$, 
\begin{equation}\label{b21}
R_i = s_i\left(1 + \alpha~p_i + \beta~p_{i+1} + \gamma~p_ip_{i+1}\right),
\end{equation}
with constants $\alpha, \beta, \gamma\in \mathbb{C}$. This operator satisfies the $(d,2,1)$-gYBE for all $\alpha, \beta, \gamma$, as seen by evaluating the two sides of the YBE
\begin{eqnarray}\label{l1}
R_iR_{i+1}R_i 
                      &=&  \left(1 + \alpha~p_{i+1} + \beta~p_i + \gamma~p_ip_{i+1}\right) \left(1 + \alpha~p_{i+2} + \beta~p_i + \gamma~p_ip_{i+2}\right) \cr
                      && \times 
                       \left(1 + \alpha~p_{i+2} + \beta~p_{i+1} + \gamma~p_{i+1}p_{i+2}\right) \left(s_is_{i+1}s_i\right), 
\end{eqnarray}
where repeated use of the permutation operator given in \eqref{sprelations3} has been applied. In a similar manner one can compute the right hand side $R_{i+1}R_iR_{i+1} $, which turns out to be equal to \eqref{l1}.  Using $p_i^2=p_i$, $p_ip_{i+1}=p_{i+1}p_i$ and $s_i^2=1$, we can show that the inverse is given by
\begin{equation}\label{b21inv}
R_i^{-1}= \left(1 - \frac{\alpha}{1+\alpha}~p_i - \frac{\beta}{1+\beta}~p_{i+1} + \frac{\alpha\beta(2+\alpha+\beta+\gamma)-\gamma}{(1+\alpha)(1+\beta)(1+\alpha+\beta+\gamma)} ~p_ip_{i+1}\right)s_i.
\end{equation} 
This expression is needed to check for which values of the parameters the $R$-matrix is unitary.

It is also easy to check that the operators in \eqref{b21} satisfy far-commutativity for braid operators, $\sigma_i\sigma_j = \sigma_j\sigma_i \; (|i-j|>1)$, by noting that
\begin{equation}
R_{i+j} = s_{i+j}\left(1 + \alpha~p_{i+j} + \beta~p_{i+j+1} + \gamma~p_{i+j}p_{i+j+1}\right)
\end{equation}
has trivial common support with the operator in \eqref{b21} for all $j>1$. 

In general, these solutions are non-unitary and generate the infinite-dimensional braid group, {\it i.e.} the image of braid group representations built using these $R$-matrices is infinite. This is seen by computing the powers of $R_i$,
\begin{equation}
R_i^n  = s_i^n\left( 1 + \alpha_{n}~p_i + \beta_{n}~p_{i+1} + \gamma_{n}~p_ip_{i+1}\right), 
\label{R_inv}
\end{equation}
where the parameters are defined recursively as
\begin{eqnarray}
\alpha_{n} & = & \alpha_1 + \beta_{n-1} + \alpha_1\beta_{n-1}, \qquad
\beta_{n}  =  \alpha_{n-1} + \beta_1 + \alpha_{n-1}\beta_1, \nonumber \\
\gamma_{n} & = & \alpha_1\alpha_{n-1} + \beta_1\beta_{n-1} + \gamma_1\gamma_{n-1} 
                       +  \gamma_1\left(1 + \alpha_{n-1} + \beta_{n-1}\right) + \gamma_{n-1}\left(1 + \alpha_1 + \beta_1\right),
                       \label{alpha_n}
\end{eqnarray} 
after identifying $\alpha_1$, $\beta_1$ and $\gamma_1$ with $\alpha$, $\beta$ and $\gamma$ in \eqref{b21}. 

By equating (\ref{R_inv}) and $R_i^\dagger$, the conditions 
\begin{eqnarray}\label{eqs_unitary}
\alpha^*  =  - \frac{\alpha}{1+\alpha}, \qquad
\beta^*  =  - \frac{\beta}{1+\beta}, \qquad
\gamma^*  =  \frac{\alpha\beta(2+\alpha+\beta+\gamma)-\gamma}{(1+\alpha)(1+\beta)(1+\alpha+\beta+\gamma)}.
\end{eqnarray}
give a family of unitary solutions that generate the infinite-dimensional braid group just as the non-unitary case. 
(\ref{eqs_unitary}) are explicitly solved by 
\begin{equation}
\alpha=e^{\mathrm{i}\theta}-1, \qquad \beta = e^{\mathrm{i}\varphi}-1, \qquad \gamma = e^{\mathrm{i}\phi}-e^{\mathrm{i}\theta}-e^{\mathrm{i}\varphi}+1
\label{sols_unitary}
\end{equation}
with $\theta$, $\varphi$ and $\phi$ angles between 0 and $2\pi$.  
The recursive definitions of the parameters of $R_i^n$ (\ref{alpha_n}) show that $R_i^n\neq 1$ for any finite $n$ when $\theta$, $\varphi$ and $\phi$ are generic. This can also be seen in the eigenvalues $\{1, \pm e^{\frac{\mathrm{i}}{2}(\theta + \varphi)}, e^{\mathrm{i}\phi}\}$ at the unitary solutions. \

There are eight real unitary solutions, four of which are shown in Table \ref{table1} in the qubit representations of \eqref{zrep1}. 
(The remaining four generate the same SLOCC classes as these four.)

\begin{table}[H]
\begin{center}
\begin{tabular}{|c|c|c|c|c|} 
\hline
& $(\alpha, \beta, \gamma)$ & $R_i$ & Eigenvalues & $n|R^n=1$ \\ \hline
1. & $(0,0,0)$ & $\tiny{\left(\begin{array}{cccc} 1 & 0 & 0 & 0 \\ 0 & 0 & 1 & 0 \\ 0 & 1 & 0 & 0 \\ 0 & 0 & 0 & 1\end{array}\right)}$ & $(-1_{(1)},1_{(3)})$ & 2 \\ \hline
2. & $(0,0,-2)$ & $\tiny{\left(\begin{array}{cccc} -1 & 0 & 0 & 0 \\ 0 & 0 & 1 & 0 \\ 0 & 1 & 0 & 0 \\ 0 & 0 & 0 & 1\end{array}\right)}$ & $(-1_{(2)},1_{(2)})$ & 2 \\ \hline
3. & $(0,-2,0)$ & $\tiny{\left(\begin{array}{cccc} -1 & 0 & 0 & 0 \\ 0 & 0 & -1 & 0 \\ 0 & 1 & 0 & 0 \\ 0 & 0 & 0 & 1\end{array}\right)}$ & $(-1,\mathrm{i},-\mathrm{i},1)$ & 4 \\ \hline
4. & $(0,-2,2)$ & $\tiny{\left(\begin{array}{cccc} 1 & 0 & 0 & 0 \\ 0 & 0 & -1 & 0 \\ 0 & 1 & 0 & 0 \\ 0 & 0 & 0 & 1\end{array}\right)}$ & $(\mathrm{i},-\mathrm{i},1_{(2)})$ & 4 \\ \hline
\end{tabular}
\caption{Unitary solutions in the 2-qubit case using operators $s_i$, $p_i$ and $p_{i+1}$. The $k$ in $a_{(k)}$ denotes the multiplicity of the eigenvalue $a$. In the last column, $n|R^n=1$ means the lowest positive integer $n$ satisfying $R^n=1$.}
\label{table1}
\end{center}
\end{table}

In the qubit representation, the $(2,2,1)$-Yang-Baxter operator takes the explicit form
\begin{equation}
R_i = \left(\begin{array}{cccc} 1+\alpha+\beta+\gamma & 0 & 0 & 0 \\ 0 & 0 & 1+\beta & 0 \\ 0 & 1+\alpha & 0 & 0 \\ 0 & 0 & 0 & 1\end{array}\right).
\label{b21matrix}
\end{equation}

The unitary operators can act as quantum gates, however not all may lead to a universal set of gates. According to a theorem by Brylinski \cite{bry} for a 2-qubit space, a gate helps building a universal set if and only if it is entangling. We can use this criterion to check which of the operators in Table \ref{table1} are entangling and can potentially lead to a universal set. 

A quantum gate is entangling if there is a vector $\ket{v_1}\otimes\ket{v_2}\in\mathbb{C}^2\otimes\mathbb{C}^2$ that gets mapped to an entangled state by the quantum gate. With this definition the gates corresponding to $(0,0,-2)$ and $(0,-2,2)$ are entangling. This assertion can be checked by seeing that these gates map the most general product state in $\mathbb{C}^2\otimes\mathbb{C}^2$, given by $a_1a_2\ket{00}+a_1b_2\ket{01} + b_1a_2\ket{10}+b_1b_2\ket{11}$, to an entangled state. For example, the operator corresponding to $(0,0,-2)$ maps this product state to $-a_1a_2\ket{00}+a_1b_2\ket{01} + b_1a_2\ket{10}+b_1b_2\ket{11}$, which is entangled. 

\paragraph{Using $s_i$ and $p_{i,i+1}$}

The operator
\begin{equation}\label{b22}
R_i = s_i\left(1 + \alpha~p_{i,i+1}\right)
\end{equation}
satisfies the $(d,2,1)$-gYBE for all values of $\alpha\in \mathbb{C}$, as can be checked by computing $R_iR_{i+1}R_i$ and $R_{i+1}R_iR_{i+1}$. 
The inverse, when $d=2$, is given by
\begin{equation}
R_i^{-1} = s_i\left(1 - \frac{\alpha}{1+2\alpha}~p_{i,i+1}\right).
\end{equation}
The image of the braid group representation built using the $(d,2,1)$-$R$-matrix in \eqref{b22} is infinite, as seen through its powers:
\begin{equation}
R_i^n  =  s_i^n\left(1 + \alpha_n~p_{i, i+1}\right), 
\end{equation}
with the parameters, when $d=2$, defined recursively as $\alpha_n  =  \alpha_1 + \alpha_{n-1} + 2\alpha_1\alpha_{n-1},$ after identifying $\alpha_1$ with $\alpha$ in \eqref{b22}.

This is unitary for $\alpha^* = -\frac{\alpha}{1+2\alpha}$, 
{\it i.e.} $\alpha=\frac12(e^{\mathrm{i}\theta}-1)$ for arbitrary angle $\theta$, which gives real solutions $\alpha=0,-1$. 
From the above recursion formula, unitary solutions with generic $\theta$ generate the infinite-dimensional braid group ($R_i^n\neq 1$ for any finite $n$).  
For the representation of the partition algebra generators in \eqref{zrep1} one obtains
\begin{equation}
R_i = \left(\begin{array}{cccc} 1 + \alpha & 0 & 0 & \alpha \\ 0 & \alpha & 1 + \alpha & 0 \\ 0 & 1 + \alpha & \alpha & 0 \\ \alpha & 0 & 0 & 1 + \alpha\end{array}\right).
\label{b22matrix}
\end{equation}
The case $\alpha=0$ is just the permutation operator $s_i$, which is not an entangler by previous considerations. That leaves us with $\alpha=-1$, which is  not an entangler either.  

\paragraph{Comparison of the unitary solutions to known cases in \cite{dye, jfranko}}
There are five families of unitary 2-qubit solutions to the Yang-Baxter equation as found in \cite{dye} and analyzed in \cite{jfranko}. The solutions in (\ref{b21}) is mapped to one of the five families whose representative is given by 
\be
\left(\begin{array}{cccc} 1 & 0 & 0 & 0 \\ 0 & 0 & \psi_1 & 0 \\ 0 & \psi_2 & 0 & 0 \\ 0 & 0 & 0 & \psi_3\end{array}\right) \qquad \mbox{with} \qquad |\psi_1|=|\psi_2|=|\psi_3|=1.
\label{dye_sol}
\ee 
Actually, the solution of the form (\ref{b21matrix}) with (\ref{sols_unitary}) becomes $\left(\begin{array}{cccc} 1 & 0 & 0 & 0 \\ 0 & 0 & e^{\mathrm{i}(\varphi-\phi)} & 0 \\ 0 & e^{\mathrm{i}(\theta-\phi)} & 0 & 0 \\ 0 & 0 & 0 & e^{-\mathrm{i}\phi}\end{array}\right)$ 
after scaling with $e^{-\mathrm{i}\phi}$, which implies that the solution belongs to the family (\ref{dye_sol}) with $\psi_1=e^{\mathrm{i}(\varphi-\phi)}$, $\psi_2=e^{\mathrm{i}(\theta-\phi)}$ and $\psi_3=e^{-\mathrm{i}\phi}$. In particular, the four real unitary solutions listed in Table \ref{table1} belong to the same family with $\psi_1=\psi_2=\psi_3=1$, $\psi_1=\psi_2=\psi_3=-1$ (after scaling with $-1$), $\psi_1=1, \psi_2=\psi_3=-1$ (after scaling with $-1$) and $\psi_1=-1, \psi_2=\psi_3=1$, respectively.  

In addition, the unitary solution in (\ref{b22}), more explicitly (\ref{b22matrix}) with $\alpha=\frac12(e^{\mathrm{i}\theta}-1)$, can also be mapped to (\ref{dye_sol}) up to the overall phase factor 
$e^{\mathrm{i}\theta}$, where $\psi_1=\psi_2=e^{-\mathrm{i}\theta}$, $\psi_3=1$, and the mapping is done by the ILO  $Q\otimes Q$ with $Q=\begin{pmatrix} 1 & 1 \\ 1 & -1\end{pmatrix}$. 

Thus, all the unitary 2-qubit solutions we obtained belong to the single family (\ref{dye_sol}) among the five described in \cite{dye, jfranko}.

\subsection{3-qubits}\label{3quR}

The number of possible operators on three sites $i$, $i+1$ and $i+2$ are $p_i$, $p_{i+1}$, $p_{i+2}$, $p_{i,i+1}$, $p_{i+1,i+2}$ and $p_{i,i+2}$, along with the permutation generators $s_i$ and $s_{i+1}$.  In order to obtain valid representations of the braid group we obtain solutions to the $(d,3,2)$-gYBE.

\paragraph{Using $s_i$, $p_i$, $p_{i+1}$ and $p_{i+2}$} 

As ansatz we propose the natural generalization of \eqref{b21} from the 2-qubit case:
\begin{equation}\label{nb31}
R_i = s_{i, i+2}\left(1 + \alpha_1~p_i +  \alpha_2~p_{i+1} + \alpha_3~p_{i+2} + \beta_1~p_ip_{i+1} + \beta_2~p_{i+1}p_{i+2} +  \beta_3~p_ip_{i+2}
+ \gamma~p_ip_{i+1}p_{i+2}\right),
\end{equation}
where $s_{i,i+2}=s_is_{i+1}s_i$ and the parameters are complex. This operator does not satisfy the $(d,3,2)$-gYBE for all values of the parameters. One can however use the identities in \eqref{sprelations1} to check that $R_iR_{i+2}R_i =R_{i+2}R_iR_{i+2}$ when 
\begin{equation}\label{nsol1}
\alpha_2 = 0, ~~ \beta_2 = -\frac{\beta_1\left(1+\alpha_3\right)}{1+\alpha_1+\beta_1},~~\gamma=\frac{\beta_1\left(\alpha_3-\alpha_1-\beta_1\right)}{1+\alpha_1+\beta_1}.
\end{equation}
The inverse is given by 
\begin{equation}\label{nb31inv}
R_i^{-1} = \left(1 + \alpha_1'~p_i  + \alpha_3'~p_{i+2} + \beta_1'~p_ip_{i+1} + \beta_2'~p_{i+1}p_{i+2} +  \beta_3'~p_ip_{i+2}
+ \gamma'~p_ip_{i+1}p_{i+2}\right)s_{i, i+2},
\end{equation}
where 
\begin{eqnarray}
\alpha_1' & = & -\frac{\alpha_1}{1+\alpha_1},~~ \alpha_3'  =  -\frac{\alpha_3}{1+\alpha_3}, \nonumber \\ 
\beta_1' & = & -\frac{\beta_1}{(1+\alpha_1)(1+\alpha_1+\beta_1)},~~\beta_2'  =  -\frac{\beta_2}{(1+\alpha_3)(1+\alpha_3+\beta_2)}, \nonumber \\ 
\beta_3' & = & \frac{\alpha_1\alpha_3(2+\alpha_1+\alpha_3)-\beta_3(1-\alpha_1\alpha_3)}{(1+\alpha_1)(1+\alpha_3)(1+\alpha_1+\alpha_3+\beta_3)},\nonumber \\ 
\gamma' & = & -\frac{\beta_1(\alpha_1-\alpha_3+\beta_1)}{(1+\alpha_1)(1+\alpha_3)(1+\alpha_1+\beta_1)}.\label{g}
\end{eqnarray}

The image of the braid group representation constructed out of \eqref{nb31} with parameters satisfying \eqref{nsol1} is infinite, as seen by computing the powers of the generalized $R$-matrix
\begin{equation}
R_i^n = s_{i, i+2}^n \left(1 + \alpha_1^{(n)}~p_i + \alpha_3^{(n)}~p_{i+2} + \beta_1^{(n)}~p_ip_{i+1} + \beta_2^{(n)}~p_{i+1}p_{i+2} + \beta_3^{(n)}~p_ip_{i+2} + \gamma^{(n)}~p_ip_{i+1}p_{i+2}\right),
\end{equation}
with the parameters defined recursively as
\begin{eqnarray}
\alpha_1^{(n)}  =  \alpha_1^{(1)} + \alpha_3^{(n-1)} + \alpha_1^{(1)}\alpha_3^{(n-1)}, \qquad
\alpha_3^{(n)}  =  \alpha_1^{(n-1)} + \alpha_3^{(1)} + \alpha_1^{(n-1)}\alpha_3^{(1)}, 
\end{eqnarray}
and
\begin{eqnarray}
\beta_1^{(n)} & =& \beta_1^{(1)} + \beta_2^{(n-1)} + \beta_1^{(1)}\beta_2^{(n-1)} + \alpha_1^{(1)}\beta_2^{(n-1)} + \alpha_3^{(n-1)}\beta_1^{(1)}, \nonumber\\ 
\beta_2^{(n)} & =& \beta_1^{(n-1)} + \beta_2^{(1)} + \beta_1^{(n-1)}\beta_2^{(1)} + \alpha_3^{(1)}\beta_1^{(n-1)} + \alpha_1^{(n-1)}\beta_2^{(1)}, \nonumber\\
\beta_3^{(n)} & = & \beta_3^{(1)} + \beta_3^{(n-1)} + \beta_3^{(1)}\beta_3^{(n-1)} + \beta_3^{(1)}\left(\alpha_1^{(n-1)} + \alpha_3^{(n-1)}\right) + \beta_3^{(n-1)}\left(\alpha_1^{(1)} + \alpha_3^{(1)}\right), \nonumber\\
\gamma^{(n)} & = & \gamma^{(1)} + \gamma^{(n-1)} + \gamma^{(1)}\gamma^{(n-1)} + \gamma^{(1)}\left(\alpha_1^{(n-1)} + \alpha_3^{(n-1)} + \beta_1^{(n-1)} + \beta_2^{(n-1)} + \beta_3^{(n-1)}\right) \nonumber \\
&  &+ \gamma^{(n-1)}\left(\alpha_1^{(1)} + \alpha_3^{(1)} + \beta_1^{(1)} + \beta_2^{(1)} + \beta_3^{(1)}\right) \nonumber \\
&  &+ \beta_1^{(n-1)}\left(\beta_1^{(1)} + \beta_3^{(1)}\right) + \beta_2^{(n-1)}\left(\beta_2^{(1)} + \beta_3^{(1)}\right) + \beta_3^{(n-1)}\left(\beta_1^{(1)} + \beta_2^{(1)}\right) \nonumber \\
&  &+ \alpha_1^{(n-1)}\beta_1^{(1)} + \alpha_3^{(n-1)}\beta_2^{(1)} + \alpha_1^{(1)}\beta_1^{(n-1)} + \alpha_3^{(1)}\beta_2^{(n-1)},
\end{eqnarray}
after identifying $\alpha_1^{(1)}$, $\alpha_3^{(1)}$, $\beta_1^{(1)}$, $\beta_2^{(1)}$, $\beta_3^{(1)}$ and $\gamma^{(1)}$ with $\alpha_1$, $\alpha_3$, $\beta_1$, $\beta_2$, $\beta_3$ and $\gamma$ in \eqref{nb31}. 

Unitary solutions occur when $\alpha_1' = \alpha_1^*$, $\alpha_3' = \alpha_3^*$, $\beta_1' = \beta_1^*$, $\beta_2' = \beta_2^*$, $\beta_3' = \beta_3^*$, and $\gamma' =\gamma^*$ with $\alpha_1',\cdots, \gamma'$ given by (\ref{g}). Their explicit form is given by 
\begin{equation}
\alpha_1=e^{\mathrm{i}\theta_1}-1,\qquad \alpha_3 = e^{\mathrm{i}\theta_3}-1,\qquad 
\beta_1=e^{\mathrm{i}\varphi_1}-e^{\mathrm{i}\theta_1}, \qquad 
\beta_3=e^{\mathrm{i}\varphi_3}-e^{\mathrm{i}\theta_1}-e^{\mathrm{i}\theta_3}+1,
\end{equation}
where $\theta_1$, $\theta_3$, $\varphi_1$ and $\varphi_3$ are arbitrary angles, and $\beta_2$ and $\gamma$ are 
obtained from (\ref{nsol1}). 
These operators with generic angles generate the infinite-dimensional braid group just as the non-unitary operators.
This is further seen from the eigenvalues at these unitary solutions given by $\{e^{-\mathrm{i}\varphi}_{(2)}, \pm e^{-\frac{\mathrm{i}}{2}\left(\theta_1 + \theta_3\right)}_{(2)}, 1_{(2)}\}$, with the $n$ in $a_{(n)}$ denoting the multiplicity of the eigenvalue $a$. 
 
 There are 16 real unitary points for the parameters in \eqref{nsol1}, of which we discuss eight. (The remaining eight fall into the same SLOCC classes as the chosen eight.)  These eight unitary solutions are not equivalent to each other and generate the GHZ, $AC-B$ and product state SLOCC classes as shown in Tables \ref{ntable1a}, \ref{ntable1b} and \ref{ntable1c}, respectively.
\begin{table}[H]
\begin{center}
\begin{tabular}{|c|c|c|c|c|} 
\hline
& $(\alpha_1, \alpha_3, \beta_1, \beta_3)$ & $R$ & Eigenvalues & $n | R^n=1$\\ \hline
1.& $(-2, -2, 2, 2)$ & $\frac{1}{2}\tiny{\left(\begin{array}{cccccccc} 0 & -1 & 1 & 0 & -1 & 0 & 0 & -1 \\ -1 & 0 & 0 & -1 & 0 & -1 & 1 & 0 \\ 1 & 0 & 0 & -1 & 0 & -1 & -1 & 0 \\ 0 & -1 & -1 & 0 & 1 & 0 & 0 & -1 \\  -1 & 0 & 0 & 1 & 0 & -1 & -1 & 0 \\ 0 & -1 & -1 & 0 & -1 & 0 & 0 & 1 \\ 0 & 1 & -1 & 0 & -1 & 0 & 0 & -1 \\ -1 & 0 & 0 & -1 & 0 & 1 & -1 & 0 \end{array}\right)}$ & $(-1_{(4)},1_{(4)})$ & 2 \\ \hline

2. & $(-2, -2, 2, 4)$ & $\frac{1}{2}\tiny{\left(\begin{array}{cccccccc} 1 & 0 & 1 & 0 & 0 & 1 & 0 & -1 \\ 0 & 1 & 0 & -1 & 1 & 0 & 1 & 0 \\ 1 & 0 & 1 & 0 & 0 & -1 & 0 & 1 \\ 0 & -1 & 0 & 1 & 1 & 0 & 1 & 0 \\  0 & 1 & 0 & 1 & 1 & 0 & -1 & 0 \\ 1 & 0 & -1 & 0 & 0 & 1 & 0 & 1 \\ 0 & 1 & 0 & 1 & -1 & 0 & 1 & 0 \\ -1 & 0 & 1 & 0 & 0 & 1 & 0 & 1 \end{array}\right)}$ & $(-1_{(2)},1_{(6)})$ & 2 \\ \hline

 3. & $(-2, 0, 2, 0)$ & $\frac{1}{2}\tiny{\left(\begin{array}{cccccccc} 0 & -1 & 0 & -1 & -1 & 0 & 1 & 0 \\ -1 & 0 & 1 & 0 & 0 & -1 & 0 & -1 \\ 0 & -1 & 0 & -1 & 1 & 0 & -1 & 0 \\ 1 & 0 & -1 & 0 & 0 & -1 & 0 & -1 \\  -1 & 0 & -1 & 0 & 0 & -1 & 0 & 1 \\ 0 & -1 & 0 & 1 & -1 & 0 & -1 & 0 \\ -1 & 0 & -1 & 0 & 0 & 1 & 0 & -1 \\ 0 & 1 & 0 & -1 & -1 & 0 & -1 & 0 \end{array}\right)}$ & $(-1_{(2)},1_{(2)},\mathrm{i}_{(2)}, -\mathrm{i}_{(2)})$ & 4 \\ \hline
 
 4. & $(-2, 0, 2, 2)$ & $\frac{1}{2}\tiny{\left(\begin{array}{cccccccc} 1 & 0 & 0 & -1 & 0 & 1 & 1 & 0 \\ 0 & 1 & 1 & 0 & 1 & 0 & 0 & -1 \\ 0 & -1 & 1 & 0 & 1 & 0 & 0 & 1 \\ 1 & 0 & 0 & 1 & 0 & -1 & 1 & 0 \\  0 & 1 & -1 & 0 & 1 & 0 & 0 & 1 \\ 1 & 0 & 0 & 1 & 0 & 1 & -1 & 0 \\ -1 & 0 & 0 & 1 & 0 & 1 & 1 & 0 \\ 0 & 1 & 1 & 0 & -1 & 0 & 0 & 1 \end{array}\right)}$ & $(\mathrm{i}_{(2)}, -\mathrm{i}_{(2)},1_{(4)})$ & 4 \\ \hline
\end{tabular}
\caption{3-qubit unitary generalized $R$-matrices generating the GHZ SLOCC class. }
\label{ntable1a}
\end{center}
\end{table}
The generalized $R$-matrices in Table \ref{ntable1a} generate the following entangled states in the GHZ SLOCC class
\begin{eqnarray}
&& \frac{1}{2}\left[-\ket{001} + \ket{010} - \ket{100} -\ket{111}\right],  \quad
 \frac{1}{2}\left[\ket{000} + \ket{010} + \ket{101} -\ket{111}\right] , \nonumber \\ 
&& \frac{1}{2}\left[-\ket{001} + \ket{011} - \ket{100} -\ket{110}\right] , \quad
 \frac{1}{2}\left[\ket{000} + \ket{011} + \ket{101} -\ket{110}\right] , 
\end{eqnarray}
which are equivalent to the standard GHZ state, $\ket{000}+\ket{111}$, by the application of appropriate ILOs, such as, for example, 
$$\left(\begin{array}{cc} a_1 & b_1\\ \mathrm{i}a_1 & -\mathrm{i}b_1\end{array}\right)\otimes\left(\begin{array}{cc} a_2 & b_2\\ -\mathrm{i}a_2 & \mathrm{i}b_2\end{array}\right)\otimes\left(\begin{array}{cc} \frac{\mathrm{i}}{4a_1a_2} & -\frac{\mathrm{i}}{4b_1b_2} \\ -\frac{1}{4a_1a_2} & -\frac{1}{4b_1b_2}\end{array}\right).$$ 

The generalized $R$-matrices in Table \ref{ntable1b} generate the partially entangled states $AC-B$ given by 
\begin{eqnarray}
 \frac{1}{2}\left[-\ket{000} - \ket{001} - \ket{100} + \ket{101}\right] ,\quad
 \frac{1}{2}\left[\ket{000} - \ket{001} + \ket{100} + \ket{101}\right], 
\end{eqnarray}
respectively. The product state SLOCC class is reported instead on Table  \ref{ntable1c}.
\begin{table}[H]
\begin{center}
\begin{tabular}{|c|c|c|c|c|} 
\hline
& $(\alpha_1, \alpha_3, \beta_1, \beta_3)$ & $R$ & Eigenvalues & $n | R^n=1$ \\ \hline

1. & $(-2, -2, 0, 2)$ & $\frac{1}{2}\tiny{\left(\begin{array}{cccccccc} -1 & -1 & 0 & 0 & -1 & 1 & 0 & 0 \\ -1 & 1 & 0 & 0 & -1 & -1 & 0 & 0 \\ 0 & 0 & -1 & -1 & 0 & 0 & -1 & 1 \\ 0 & 0 & -1 & 1 & 0 & 0 & -1 & -1 \\  -1 & -1 & 0 & 0 & 1 & -1 & 0 & 0 \\ 1 & -1 & 0 & 0 & -1 & -1 & 0 & 0 \\ 0 & 0 & -1 & -1 & 0 & 0 & 1 & -1 \\ 0 & 0 & 1 & -1 & 0 & 0 & -1 & -1 \end{array}\right)}
$ & $(-1_{(4)},1_{(4)})$ & 2 \\ \hline

2. & $(-2, 0, 0, 2)$ & $\frac{1}{2}\tiny{\left(\begin{array}{cccccccc} 1 & 1 & 0 & 0 & -1 & 1 & 0 & 0 \\ -1 & 1 & 0 & 0 & 1 & 1 & 0 & 0 \\ 0 & 0 & 1 & 1 & 0 & 0 & -1 & 1 \\ 0 & 0 & -1 & 1 & 0 & 0 & 1 & 1 \\  1 & 1 & 0 & 0 & 1 & -1 & 0 & 0 \\ 1 & -1 & 0 & 0 & 1 & 1 & 0 & 0 \\ 0 & 0 & 1 & 1 & 0 & 0 & 1 & -1 \\ 0 & 0 & 1 & -1 & 0 & 0 & 1 & 1 \end{array}\right)}
$ & $(\mathrm{i}_{(2)},-\mathrm{i}_{(2)},1_{(4)})$ & 4 \\ \hline

\end{tabular}
\caption{3-qubit unitary generalized $R$-matrices generating the $AC-B$ SLOCC class. }
\label{ntable1b}
\end{center}
\end{table}
\begin{table}[H]
\begin{center}
\begin{tabular}{|c|c|c|c|c|} 
\hline
& $(\alpha_1, \alpha_3, \beta_1, \beta_3)$ & $R$ & Eigenvalues & $n | R^n=1$ \\ \hline

1. &  $(-2, -2, 0, 4)$ & $
\tiny{\left(\begin{array}{cccccccc} 0 & 0 & 0 & 0 & 0 & 1 & 0 & 0 \\ 0 & 1 & 0 & 0 & 0 & 0 & 0 & 0 \\ 0 & 0 & 0 & 0 & 0 & 0 & 0 & 1 \\ 0 & 0 & 0 & 1 & 0 & 0 & 0 & 0 \\  0 & 0 & 0 & 0 & 1 & 0 & 0 & 0 \\ 1 & 0 & 0 & 0 & 0 & 0 & 0 & 0 \\ 0 & 0 & 0 & 0 & 0 & 0 & 1 & 0 \\ 0 & 0 & 1 & 0 & 0 & 0 & 0 & 0 \end{array}\right)
}
$ & $(-1_{(2)},1_{(6)})$ & 2 \\ \hline

2. & $(-2, 0, 0, 0)$ & $
\tiny{\left(\begin{array}{cccccccc} 0 & 0 & 0 & 0 & -1 & 0 & 0 & 0 \\ -1 & 0 & 0 & 0 & 0 & 0 & 0 & 0 \\ 0 & 0 & 0 & 0 & 0 & 0 & -1 & 0 \\ 0 & 0 & -1 & 0 & 0 & 0 & 0 & 0 \\  0 & 0 & 0 & 0 & 0 & -1 & 0 & 0 \\ 0 & -1 & 0 & 0 & 0 & 0 & 0 & 0 \\ 0 & 0 & 0 & 0 & 0 & 0 & 0 & -1 \\ 0 & 0 & 0 & -1 & 0 & 0 & 0 & 0 \end{array}\right)}
$ & $(\mathrm{i}_{(2)},-\mathrm{i}_{(2)},-1_{(2)},1_{(2)})$ & 4 \\ \hline

\end{tabular}
\caption{3-qubit unitary generalized $R$-matrices generating the product state SLOCC class. }
\label{ntable1c}
\end{center}
\end{table}

\paragraph{Using $s_i$, $p_{i,i+1}$ and $p_{i+1,i+2}$}

The ansatz
\begin{equation}\label{nb32}
R_i = s_{i, i+2}\left(1 + \alpha~p_{i,i+1} + \beta~p_{i+1,i+2} + \gamma~p_{i,i+1}p_{i+1,i+2} + \delta~p_{i,i+2}\right),
\end{equation}
with $s_{i,i+2} = s_is_{i+1}s_i$, satisfies the $(d,3,2)$-gYBE when $\gamma=-\frac{\alpha+\beta}{2}$. This is seen after simplifying the expressions in the $(d,3,2)$-gYBE using the swapping property of the permutation  operator in \eqref{sprelations4}, as done before.

The inverse, when $d=2$ and $\gamma=-\frac{\alpha+\beta}{2}$ is given by
\begin{equation}\label{nb32bis}
R_i^{-1} = \left(1 + \alpha'~p_{i,i+1} + \beta'~p_{i+1,i+2} + \gamma'~p_{i,i+1}p_{i+1,i+2} + \delta'~p_{i,i+2}\right)s_{i, i+2},
\end{equation}
with 
\begin{eqnarray}
\alpha'  =  -\frac{\alpha}{1+2\alpha},\quad \beta'  =  -\frac{\beta}{1+2\beta}, \quad
\delta'  =  -\frac{\delta}{1+2\delta},\quad \gamma' = \frac{\alpha+\beta+4\alpha\beta}{2+4(\alpha+\beta+2\alpha\beta)},
\end{eqnarray}
 which results in a unitary family when $\alpha'=\alpha^*$, $\beta'=\beta^*$, $\gamma'=\gamma^*$, and $\delta'=\delta^*$ that are solved by 
\begin{equation}
\alpha=\frac12(e^{\mathrm{i}\theta}-1),\qquad \beta=\frac12(e^{\mathrm{i}\varphi}-1),\qquad \delta=\frac12(e^{\mathrm{i}\phi}-1)
\end{equation}
with $\gamma=-\frac{\alpha+\beta}{2}$ for $\theta$, $\varphi$ and $\phi$ arbitrary angles.  The eigenvalues of this operator at the unitary family are given by $\{e^{\mathrm{i}\phi}_{(4)}, \pm e^{\frac{\mathrm{i}}{2}\left(\theta + \varphi\right)}_{(2)}\}$. 
 
 There are eight real unitary solutions, out of which four inequivalent unitary solutions are shown in Table \ref{ntable2}. 
 \begin{table}[H]
\begin{center}
\begin{tabular}{|c|c|c|c|c|} 
\hline
& $(\alpha, \beta, \delta)$ & $R_i$ & Eigenvalues & $n | R^n=1$ \\ \hline

1.&  $(-1,-1,-1)$ & $\frac{1}{2}\tiny{\left(\begin{array}{cccccccc} -1 & 0 & 0 & 0 & 0 & 0 & 0 & 0 \\ 0 & 0 & 0 & 0 & -1 & 0 & 0 & 0 \\ 0 & 0 & -1 & 0 & 0 & 0 & 0 & 0 \\ 0 & 0 & 0 & 0 & 0 & 0 & -1 & 0 \\  0 & -1 & 0 & 0 & 0 & 0 & 0 & 0 \\ 0 & 0 & 0 & 0 & 0 & -1 & 0 & 0 \\ 0 & 0 & 0 & -1 & 0 & 0 & 0 & 0 \\ 0 & 0 & 0 & 0 & 0 & 0 & 0 & -1 \end{array}\right)}$ & $(-1_{(6)},1_{(2)})$ & 2 \\ \hline

2. & $(-1,-1,0)$ & $\tiny{\left(\begin{array}{cccccccc} 0 & 0 & 0 & 0 & 0 & 1 & 0 & 0 \\ 0 & 1 & 0 & 0 & 0 & 0 & 0 & 0 \\ 0 & 0 & 0 & 0 & 0 & 0 & 0 & 1 \\ 0 & 0 & 0 & 1 & 0 & 0 & 0 & 0 \\  0 & 0 & 0 & 0 & 1 & 0 & 0 & 0 \\ 1 & 0 & 0 & 0 & 0 & 0 & 0 & 0 \\ 0 & 0 & 0 & 0 & 0 & 0 & 1 & 0 \\ 0 & 0 & 1 & 0 & 0 & 0 & 0 & 0 \end{array}\right)}$ & $(-1_{(2)},1_{(6)})$ & 2 \\ \hline

3. & $(-1,0,-1)$ & $\frac{1}{2}\tiny{\left(\begin{array}{cccccccc} -1 & 0 & 0 & 1 & 0 & -1 & -1 & 0 \\ 0 & -1 & -1 & 0 & -1 & 0 & 0 & 1 \\ 0 & 1 & -1 & 0 & -1 & 0 & 0 & -1 \\ -1 & 0 & 0 & -1 & 0 & 1 & -1 & 0 \\  0 & -1 & 1 & 0 & -1 & 0 & 0 & -1 \\ -1 & 0 & 0 & -1 & 0 & -1 & 1 & 0 \\ 1 & 0 & 0 & -1 & 0 & -1 & -1 & 0 \\ 0 & -1 & -1 & 0 & 1 & 0 & 0 & -1 \end{array}\right)}$ & $(-1_{(4)},\mathrm{i}_{(2)},-\mathrm{i}_{(2)})$ & 4 \\ \hline
 
4. & $(-1,0,0)$ & $\frac{1}{2}\tiny{\left(\begin{array}{cccccccc} 1 & 0 & 0 & 1 & 0 & 1 & -1 & 0 \\ 0 & 1 & -1 & 0 & 1 & 0 & 0 & 1 \\ 0 & 1 & 1 & 0 & -1 & 0 & 0 & 1 \\ -1 & 0 & 0 & 1 & 0 & 1 & 1 & 0 \\  0 & 1 & 1 & 0 & 1 & 0 & 0 & -1 \\ 1 & 0 & 0 & -1 & 0 & 1 & 1 & 0 \\ 1 & 0 & 0 & 1 & 0 & -1 & 1 & 0 \\ 0 & -1 & 1 & 0 & 1 & 0 & 0 & 1 \end{array}\right)}$ & $(\mathrm{i}_{(2)},-\mathrm{i}_{(2)},1_{(4)})$ & 4 \\ \hline
\end{tabular}
\caption{Unitary solutions in the 3-qubit case for the generalized Yang--Baxter operator in \eqref{nb32} with $\gamma=-\frac{\alpha+\beta}{2}$. }
\label{ntable2}
\end{center}
\end{table}
 
The image of the braid group representations constructed out of the generalized $R$-matrix in \eqref{nb32} for $d=2$ and $\gamma=-\frac{\alpha+\beta}{2}$, for both the unitary and non-unitary cases, is once again infinite as in the previous cases, as seen by computing the powers of the generalized $R$-matrix,
 \begin{equation}
 R_i^n = s_{i, i+2}^n\left(1 + \alpha_n~p_{i, i+1} + \beta_n~p_{i+1, i+2} + \gamma_n~p_{i, i+1}p_{i+1, i+2} + \delta_n~p_{i, i+2}\right),
 \end{equation}
 with the parameters defined recursively as
 \begin{eqnarray}
 \alpha_n & = & \alpha_1 + \beta_{n-1} + 2\alpha_1\beta_{n-1}, \qquad
 \beta_n  =  \alpha_{n-1} + \beta_1 + 2\alpha_{n-1}\beta_1, \cr
 \gamma_n & = & \gamma_1 + \gamma_{n-1} + 4\gamma_1\gamma_{n-1} + 2\delta_1\gamma_{n-1} + 2\gamma_1\delta_{n-1} + \delta_{n-1}\left(\alpha_1 + \beta_1\right) + \delta_1\left(\alpha_{n-1} + \beta_{n-1}\right) \nonumber \\
 &  &+ 2\gamma_{n-1}\left(\alpha_1 + \beta_1\right) + 2\gamma_1\left(\alpha_{n-1} + \beta_{n-1}\right) + \alpha_1\alpha_{n-1} + \beta_1\beta_{n-1}, \cr
 \delta_n & = & \delta_1 + \delta_{n-1} + 2\delta_1\delta_{n-1}, 
 \end{eqnarray}
 after identifying $\alpha_1$, $\beta_1$, $\gamma_1$ and $\delta_1$ with $\alpha$, $\beta$, $\gamma$ and $\delta$ in \eqref{nb32}.
 
 The generalized $R$-matrices in the first two rows of Table \ref{ntable2} are not entanglers, as they generate just the product state SLOCC class. The generalized $R$-matrices of the second two rows, on the other hand, are both entanglers generating the GHZ SLOCC class. In particular the states they generate by acting on $\ket{000}$ are
 \begin{eqnarray}
 \frac{1}{2}\left[-\ket{000}-\ket{011}-\ket{101}+\ket{110}\right],\qquad
 \frac{1}{2}\left[\ket{000}-\ket{011}+\ket{101}+\ket{110}\right],
 \end{eqnarray}
 which are equivalent to the standard GHZ state $\ket{000}+\ket{111}$ by appropriate ILOs, such as, for example,
 $$\left(\begin{array}{cc} a_1 & b_1\\ \mathrm{i}a_1 & -\mathrm{i}b_1\end{array}\right)\otimes\left(\begin{array}{cc} a_2 & b_2\\ \mathrm{i}a_2 & -\mathrm{i}b_2\end{array}\right)\otimes\left(\begin{array}{cc} -\frac{1}{4a_1a_2} & -\frac{1}{4b_1b_2} \\ \frac{\mathrm{i}}{4a_1a_2} & -\frac{\mathrm{i}}{4b_1b_2}\end{array}\right).$$
 
\subsection{4-qubits}\label{4quR}

As one increases the number of qubits, the analytic computation for the generalized $R$-matrices gets more tedious. To illustrate the feasibility of the method, we write down the answers for the 4-qubit case as well. We use the operators $s_i$, $p_i$, $p_{i+1}$, $p_{i+2}$ and $p_{i+3}$ to build the generalized $R$-matrices. The generalized $R$-matrices that satisfy the $(d,4,2)$-gYBE and the $(d,4,3)$-gYBE also satisfy far-commutativity yielding braid group representations. These are the cases we  focus on.

\subsubsection*{The $(d,4,2)$-generalized $R$-matrix} 

The generalized $R$-matrix 
\begin{eqnarray} \label{nb41}
R_i & = & s_{i, i+2}\left( 1 + \alpha_1~p_i + \alpha_3~p_{i+2} + \beta_1~p_ip_{i+1} + \beta_2~p_ip_{i+2}\right. \nonumber \\
      &  & + \beta_3~p_ip_{i+3} + \beta_4~p_{i+1}p_{i+2} + \beta_6~p_{i+2}p_{i+3}+ \gamma_1~p_ip_{i+1}p_{i+2} +  \nonumber \\
      &  &+ \left.  \gamma_2~p_ip_{i+1}p_{i+3} + \gamma_3~p_ip_{i+2}p_{i+3} + \gamma_4~p_{i+1}p_{i+2}p_{i+3} +  \delta~p_ip_{i+1}p_{i+2}p_{i+3}\right),
\end{eqnarray}      
 with $s_{i, i+2}=s_is_{i+1}s_i$, satisfies the $(d,4,2)$-gYBE for 
 \begin{eqnarray}
 \beta_4 & = & -\frac{\beta_1\left(1+\alpha_3\right)}{1+\alpha_1+\beta_1},~~ \beta_6  =  -\frac{\beta_3\left(1+\alpha_3\right)}{1+\alpha_1+\beta_3}, \nonumber \\
 \gamma_1 & = & -\frac{\beta_1\left(\alpha_1+\beta_1-\alpha_3\right)}{1+\alpha_1+\beta_1},~~ \gamma_3  =  -\frac{\beta_3\left(\alpha_1+\beta_3-\alpha_3\right)}{1+\alpha_1+\beta_3}, \nonumber \\
 \gamma_4 & = & \left(1+\alpha_3\right)\left[\frac{\beta_1}{1+\alpha_1+\beta_1} - \frac{\left(1+\alpha_1\right)\left(\beta_1+\gamma_2\right)}{\left(1+\alpha_1+\beta_3\right)\left(1+\alpha_1+\beta_1+\beta_3+\gamma_2\right)}\right], \nonumber \\
 \delta & = & -\gamma_2 + \frac{\left(1+\alpha_3\right)\left[-\beta_1\beta_3\left(2\alpha_1+\beta_1+\beta_3+2\right)+\gamma_2\left(\left(1+\alpha_1\right)^2-\beta_1\beta_3\right)\right]}{\left(1+\alpha_1+\beta_1\right)\left(1+\alpha_1+\beta_3\right)\left(1+\alpha_1+\beta_1+\beta_3+\gamma_2\right)}.
 \end{eqnarray}    
 The solutions become unitary when 
 \begin{eqnarray}
\alpha_1^* & = & -\frac{\alpha_1}{1+\alpha_1},~~ \alpha_3^*  =  -\frac{\alpha_3}{1+\alpha_3}, \nonumber \\ 
\beta_1^* & = & -\frac{\beta_1}{(1+\alpha_1)(1+\alpha_1+\beta_1)},~~\beta_3^*  =  -\frac{\beta_3}{(1+\alpha_1)(1+\alpha_1+\beta_3)}, \nonumber \\ 
\beta_2^* & = & \frac{\alpha_1}{1+\alpha_1} - \frac{1}{1+\alpha_3} + \frac{1}{1+\alpha_1 +\alpha_3+\beta_2},  \nonumber \\
\gamma_2^* & = & \frac{\beta_1}{(1+\alpha_1)(1+\alpha_1+\beta_1)} - \frac{1}{1+\alpha_1+\beta_3}+\frac{1}{1+\alpha_1+\beta_1+\beta_3+\gamma_2},\label{d42}
\end{eqnarray} 
which is solved by 
\begin{eqnarray}
\alpha_1 & = & e^{\mathrm{i}\theta_1} -1, ~~ \alpha_3 = e^{\mathrm{i}\theta_3} -1, \nonumber \\ 
\beta_1 & = & e^{\mathrm{i}\phi_1} -e^{\mathrm{i}\theta_1}, ~~ \beta_3 = e^{\mathrm{i}\phi_3} -e^{\mathrm{i}\theta_1},\nonumber \\ 
\beta_2 & = & e^{\mathrm{i}\phi_2} - e^{\mathrm{i}\theta_1} - e^{\mathrm{i}\theta_3} + 1,~~ \gamma_2 = e^{\mathrm{i}\varphi_2} - e^{\mathrm{i}\phi_1} -e^{\mathrm{i}\phi_3} + e^{\mathrm{i}\theta_1}, \label{uc42}  
\end{eqnarray}
for arbitrary angles $\theta_1$, $\theta_3$, $\phi_1$, $\phi_3$ and $\varphi_2$. The eigenvalues at these unitary solutions are given by $\{1_{(4)}, \pm e^{\frac{\mathrm{i}}{2}(\theta_1 + \theta_3)}_{(4)}, e^{\mathrm{i}\phi_2}_{(4)}\}$. 

There are 64 real unitary generalized $R$-matrices in this case. They encompass four sets of eigenvalues, with 16 unitary generalized $R$-matrices in each of these sets. The unitary generalized $R$-matrices are inequivalent when they belong to different eigenvalue sets, however when they belong to the same eigenvalue set they may or may not be equivalent. We write down one unitary solution from each set of eigenvalues. 
 
\paragraph{Eigenvalues $\in \{-1_{(8)},1_{(8)}\}$}
The solution with $\left(\alpha_1, \alpha_3, \beta_1, \beta_2, \beta_3, \gamma_2\right) = \left(-2,-2,0,2,0,0\right)$ reads explicitly
\begin{equation}\label{d42_1}
\frac{1}{2}\left(\tiny{\begin{array}{cccccccc|cccccccc} -1 & 0 & -1 & 0 & 0 & 0 & 0 & 0 & -1 & 0 & 1 & 0 & 0 & 0 & 0 & 0 \\ 0 & -1 & 0 & -1 & 0 & 0 & 0 & 0 & 0 & -1 & 0 & 1 & 0 & 0 & 0 & 0 \\ -1 & 0 & 1 & 0 & 0 & 0 & 0 & 0 & -1 & 0 & -1 & 0 & 0 & 0 & 0 & 0 \\ 0 & -1 & 0 & 1 & 0 & 0 & 0 & 0 & 0 & -1 & 0 & -1 & 0 & 0 & 0 & 0 \\ 0 & 0 & 0 & 0 & -1 & 0 & -1 & 0 & 0 & 0 & 0 & 0 & -1 & 0 & 1 & 0 \\ 0 & 0 & 0 & 0 & 0 & -1 & 0 & -1 & 0 & 0 & 0 & 0 & 0 & -1 & 0 & 1 \\ 0 & 0 & 0 & 0 & -1 & 0 & 1 & 0 & 0 & 0 & 0 & 0 & -1 & 0 & -1 & 0 \\ 0 & 0 & 0 & 0 & 0 & -1 & 0 & 1 & 0 & 0 & 0 & 0 & 0 & -1 & 0 & -1 \\ \hline -1 & 0 & -1 & 0 & 0 & 0 & 0 & 0 & 1 & 0 & -1 & 0 & 0 & 0 & 0 & 0 \\ 0 & -1 & 0 & -1 & 0 & 0 & 0 & 0 & 0 & 1 & 0 & -1 & 0 & 0 & 0 & 0 \\ 1 & 0 & -1 & 0 & 0 & 0 & 0 & 0 & -1 & 0 & -1 & 0 & 0 & 0 & 0 & 0 \\ 0 & 1 & 0 & -1 & 0 & 0 & 0 & 0 & 0 & -1 & 0 & -1 & 0 & 0 & 0 & 0 \\ 0 & 0 & 0 & 0 & -1 & 0 & -1 & 0 & 0 & 0 & 0 & 0 & 1 & 0 & -1 & 0 \\ 0 & 0 & 0 & 0 & 0 & -1 & 0 & -1 & 0 & 0 & 0 & 0 & 0 & 1 & 0 & -1 \\ 0 & 0 & 0 & 0 & 1 & 0 & -1 & 0 & 0 & 0 & 0 & 0 & -1 & 0 & -1 & 0 \\ 0 & 0 & 0 & 0 & 0 & 1 & 0 & -1 & 0 & 0 & 0 & 0 & 0 & -1 & 0 & -1  \end{array}}\right).
\end{equation}
This generates $ \frac{1}{2}\left[-\ket{0000} - \ket{0010} - \ket{1000} + \ket{1010}\right]$ from $\ket{0000}$.
 
\paragraph{Eigenvalues $\in \{-1_{(4)},1_{(12)}\}$}
The solution for $\left(\alpha_1, \alpha_3, \beta_1, \beta_2, \beta_3, \gamma_2\right) = \left(-2,-2,0,4,0,2\right)$ reads
\begin{equation}
\frac{1}{4}\left(\tiny{\begin{array}{cccccccc|cccccccc} 1 & 1 & 0 & 0 & 1 & 1 & 0 & 0 & 0 & 0 & 3 & -1 & 0 & 0 & -1 & -1 \\ 1 & 1 & 0 & 0 & 1 & 1 & 0 & 0 & 0 & 0 & -1 & 3 & 0 & 0 & -1 & -1 \\ 0 & 0 & 3 & -1 & 0 & 0 & -1 & -1 & 1 & 1 & 0 & 0 & 1 & 1 & 0 & 0 \\ 0 & 0 & -1 & 3 & 0 & 0 & -1 & -1 & 1 & 1 & 0 & 0 & 1 & 1 & 0 & 0 \\ 1 & 1 & 0 & 0 & 1 & 1 & 0 & 0 & 0 & 0 & -1 & -1 & 0 & 0 & 3 & -1 \\ 1 & 1 & 0 & 0 & 1 & 1 & 0 & 0 & 0 & 0 & -1 & -1 & 0 & 0 & -1 & 3 \\ 0 & 0 & -1 & -1 & 0 & 0 & 3 & -1 & 1 & 1 & 0 & 0 & 1 & 1 & 0 & 0 \\ 0 & 0 & -1 & -1 & 0 & 0 & -1 & 3 & 1 & 1 & 0 & 0 & 1 & 1 & 0 & 0 \\ \hline 0 & 0 & 1 & 1 & 0 & 0 & 1 & 1 & 3 & -1 & 0 & 0 & -1 & -1 & 0 & 0 \\ 0 & 0 & 1 & 1 & 0 & 0 & 1 & 1 & -1 & 3 & 0 & 0 & -1 & -1 & 0 & 0 \\ 3 & -1 & 0 & 0 & -1 & -1 & 0 & 0 & 0 & 0 & 1 & 1 & 0 & 0 & 1 & 1 \\ -1 & 3 & 0 & 0 & -1 & -1 & 0 & 0 & 0 & 0 & 1 & 1 & 0 & 0 & 1 & 1 \\ 0 & 0 & 1 & 1 & 0 & 0 & 1 & 1 & -1 & -1 & 0 & 0 & 3 & -1 & 0 & 0 \\ 0 & 0 & 1 & 1 & 0 & 0 & 1 & 1 & -1 & -1 & 0 & 0 & -1 & 3 & 0 & 0 \\ -1 & -1 & 0 & 0 & 3 & -1 & 0 & 0 & 0 & 0 & 1 & 1 & 0 & 0 & 1 & 1 \\ -1 & -1 & 0 & 0 & -1 & 3 & 0 & 0 & 0 & 0 & 1 & 1 & 0 & 0 & 1 & 1  \end{array}}\right).
\end{equation}
This generates $ \frac{1}{4}\left[\ket{0000} + \ket{0001} + \ket{0100} + \ket{0101} + 3\ket{1010} - \ket{1011} - \ket{1110} - \ket{1111}\right]$ from $\ket{0000}$.

\paragraph{Eigenvalues $\in \{-\mathrm{i}_{(4)}, \mathrm{i}_{(4)},1_{(8)}\}$}

The solution with $\left(\alpha_1, \alpha_3, \beta_1, \beta_2, \beta_3, \gamma_2\right) = \left(-2,0,0,2,2,0\right)$ reads
\begin{equation}
\frac{1}{2}
 \left(\tiny{\begin{array}{cccccccc|cccccccc} 1 & 0 & 0 & -1 & 0 & 0 & 0 & 0 & 0 & 1 & 1 & 0 & 0 & 0 & 0 & 0 \\ 0 & 1 & -1 & 0 & 0 & 0 & 0 & 0 & 1 & 0 & 0 & 1 & 0 & 0 & 0 & 0 \\ 0 & 1 & 1 & 0 & 0 & 0 & 0 & 0 & 1 & 0 & 0 & -1 & 0 & 0 & 0 & 0 \\ 1 & 0 & 0 & 1 & 0 & 0 & 0 & 0 & 0 & 1 & -1 & 0 & 0 & 0 & 0 & 0 \\ 0 & 0 & 0 & 0 & 1 & 0 & 0 & -1 & 0 & 0 & 0 & 0 & 0 & 1 & 1 & 0 \\ 0 & 0 & 0 & 0 & 0 & 1 & -1 & 0 & 0 & 0 & 0 & 0 & 1 & 0 & 0 & 1 \\ 0 & 0 & 0 & 0 & 0 & 1 & 1 & 0 & 0 & 0 & 0 & 0 & 1 & 0 & 0 & -1 \\ 0 & 0 & 0 & 0 & 1 & 0 & 0 & 1 & 0 & 0 & 0 & 0 & 0 & 1 & -1 & 0 \\ \hline 0 & -1 & 1 & 0 & 0 & 0 & 0 & 0 & 1 & 0 & 0 & 1 & 0 & 0 & 0 & 0 \\ -1 & 0 & 0 & 1 & 0 & 0 & 0 & 0 & 0 & 1 & 1 & 0 & 0 & 0 & 0 & 0 \\ 1 & 0 & 0 & 1 & 0 & 0 & 0 & 0 & 0 & -1 & 1 & 0 & 0 & 0 & 0 & 0 \\ 0 & 1 & 1 & 0 & 0 & 0 & 0 & 0 & -1 & 0 & 0 & 1 & 0 & 0 & 0 & 0 \\ 0 & 0 & 0 & 0 & 0 & -1 & 1 & 0 & 0 & 0 & 0 & 0 & 1 & 0 & 0 & 1 \\ 0 & 0 & 0 & 0 & -1 & 0 & 0 & 1 & 0 & 0 & 0 & 0 & 0 & 1 & 1 & 0 \\ 0 & 0 & 0 & 0 & 1 & 0 & 0 & 1 & 0 & 0 & 0 & 0 & 0 & -1 & 1 & 0 \\ 0 & 0 & 0 & 0 & 0 & 1 & 1 & 0 & 0 & 0 & 0 & 0 & -1 & 0 & 0 & 1  \end{array}}\right)
.
\end{equation}
This generates $ \frac{1}{2}\left[\ket{0000} + \ket{0011} - \ket{1001} + \ket{1010}\right]$ from $\ket{0000}$.

\paragraph{Eigenvalues $\in \{-\mathrm{i}_{(4)}, \mathrm{i}_{(4)},-1_{(4)}, 1_{(4)}\}$}
The solution with $\left(\alpha_1, \alpha_3, \beta_1, \beta_2, \beta_3, \gamma_2\right) = \left(-2,0,0,0,2,0\right)$ reads
\begin{equation}\label{d42_4}
\frac{1}{2}\left(\tiny{\begin{array}{cccccccc|cccccccc} 0 & 0 & -1 & -1 & 0 & 0 & 0 & 0 & -1 & 1 & 0 & 0 & 0 & 0 & 0 & 0 \\ 0 & 0 & -1 & -1 & 0 & 0 & 0 & 0 & 1 & -1 & 0 & 0 & 0 & 0 & 0 & 0 \\ -1 & 1 & 0 & 0 & 0 & 0 & 0 & 0 & 0 & 0 & -1 & -1 & 0 & 0 & 0 & 0 \\ 1 & -1 & 0 & 0 & 0 & 0 & 0 & 0 & 0 & 0 & -1 & -1 & 0 & 0 & 0 & 0 \\ 0 & 0 & 0 & 0 & 0 & 0 & -1 & -1 & 0 & 0 & 0 & 0 & -1 & 1 & 0 & 0 \\ 0 & 0 & 0 & 0 & 0 & 0 & -1 & -1 & 0 & 0 & 0 & 0 & 1 & -1 & 0 & 0 \\ 0 & 0 & 0 & 0 & -1 & 1 & 0 & 0 & 0 & 0 & 0 & 0 & 0 & 0 & -1 & -1 \\ 0 & 0 & 0 & 0 & 1 & -1 & 0 & 0 & 0 & 0 & 0 & 0 & 0 & 0 & -1 & -1 \\ \hline -1 & -1 & 0 & 0 & 0 & 0 & 0 & 0 & 0 & 0 & -1 & 1 & 0 & 0 & 0 & 0 \\ -1 & -1 & 0 & 0 & 0 & 0 & 0 & 0 & 0 & 0 & 1 & -1 & 0 & 0 & 0 & 0 \\ 0 & 0 & -1 & 1 & 0 & 0 & 0 & 0 & -1 & -1 & 0 & 0 & 0 & 0 & 0 & 0 \\ 0 & 0 & 1 & -1 & 0 & 0 & 0 & 0 & -1 & -1 & 0 & 0 & 0 & 0 & 0 & 0 \\ 0 & 0 & 0 & 0 & -1 & -1 & 0 & 0 & 0 & 0 & 0 & 0 & 0 & 0 & -1 & 1 \\ 0 & 0 & 0 & 0 & -1 & -1 & 0 & 0 & 0 & 0 & 0 & 0 & 0 & 0 & 1 & -1 \\ 0 & 0 & 0 & 0 & 0 & 0 & -1 & 1 & 0 & 0 & 0 & 0 & -1 & -1 & 0 & 0 \\ 0 & 0 & 0 & 0 & 0 & 0 & 1 & -1 & 0 & 0 & 0 & 0 & -1 & -1 & 0 & 0  \end{array}}\right).
\end{equation} 
 This generates $ \frac{1}{2}\left[-\ket{0010} + \ket{0011} - \ket{1000} - \ket{1001}\right]$
from $\ket{0000}$.
 
\subsubsection*{The $(d,4,3)$-generalized $R$-matrix} 

The generalized $R$-matrix 
\begin{eqnarray} \label{nb42}
R_i & = & s_{i, i+3}\left( 1 + \alpha_1~p_i + \alpha_4~p_{i+3} + \beta_1~p_ip_{i+1} \right. \nonumber \\
      &  &+ \left. \beta_2~p_ip_{i+2} + \beta_3~p_ip_{i+3} + \beta_5~p_{i+1}p_{i+3} + \beta_6~p_{i+2}p_{i+3} \right. \nonumber \\
      &  &+ \left. \gamma_1~p_ip_{i+1}p_{i+2} + \gamma_2~p_ip_{i+1}p_{i+3} + \gamma_3~p_ip_{i+2}p_{i+3} + \gamma_4~p_{i+1}p_{i+2}p_{i+3} \right. \nonumber \\
      &  &+ \left. \delta~p_ip_{i+1}p_{i+2}p_{i+3}\right),
\end{eqnarray}      
 with $s_{i, i+3}=s_{i+2}s_{i+1}s_is_{i+1}s_{i+2}$, satisfies the $(d,4,3)$-gYBE for 
 \begin{eqnarray}
 \beta_5 & = & -\frac{\beta_1\left(1+\alpha_4\right)}{1+\alpha_1+\beta_1},~~ \beta_6  =  -\frac{\beta_2\left(1+\alpha_4\right)}{1+\alpha_1+\beta_2}, \nonumber \\
 \gamma_2 & = & -\frac{\beta_1\left(\alpha_1+\beta_1-\alpha_4\right)}{1+\alpha_1+\beta_1},~~ \gamma_3  =  -\frac{\beta_2\left(\alpha_1+\beta_2-\alpha_4\right)}{1+\alpha_1+\beta_2}, \nonumber \\
 \gamma_4 & = & \left(1+\alpha_4\right)\left[\frac{\beta_1}{1+\alpha_1+\beta_1} - \frac{\left(1+\alpha_1\right)\left(\beta_1+\gamma_1\right)}{\left(1+\alpha_1+\beta_2\right)\left(1+\alpha_1+\beta_1+\beta_2+\gamma_1\right)}\right], \nonumber \\
 \delta & = & -\gamma_1 + \frac{\left(1+\alpha_4\right)\left[-\beta_1\beta_2\left(2\alpha_1+\beta_1+\beta_2+2\right)+\gamma_1\left(\left(1+\alpha_1\right)^2-\beta_1\beta_2\right)\right]}{\left(1+\alpha_1+\beta_1\right)\left(1+\alpha_1+\beta_2\right)\left(1+\alpha_1+\beta_1+\beta_2+\gamma_1\right)}.
 \end{eqnarray} 
 The solutions become unitary when 
 \begin{eqnarray}
\alpha_1^* & = & -\frac{\alpha_1}{1+\alpha_1},~~ \alpha_4^*  =  -\frac{\alpha_4}{1+\alpha_4}, \nonumber \\ 
\beta_1^* & = & -\frac{\beta_1}{(1+\alpha_1)(1+\alpha_1+\beta_1)},~~\beta_2^*  =  -\frac{\beta_2}{(1+\alpha_1)(1+\alpha_1+\beta_2)}, \nonumber \\ 
\beta_3^* & = & \frac{\alpha_1}{1+\alpha_1} - \frac{1}{1+\alpha_4} + \frac{1}{1+\alpha_1 +\alpha_4+\beta_3}, \nonumber \\ 
\gamma_1^* & = & \frac{\beta_1}{(1+\alpha_1)(1+\alpha_1+\beta_1)} - \frac{1}{1+\alpha_1+\beta_2}+\frac{1}{1+\alpha_1+\beta_1+\beta_2+\gamma_1},\label{d43}
\end{eqnarray} 
which is solved by 
\begin{eqnarray}
\alpha_1 & = & e^{\mathrm{i}\theta_1} -1, ~~ \alpha_4 = e^{\mathrm{i}\theta_4} -1, \nonumber \\ 
\beta_1 & = & e^{\mathrm{i}\phi_1} -e^{\mathrm{i}\theta_1}, ~~ \beta_2 = e^{\mathrm{i}\phi_2} -e^{\mathrm{i}\theta_1}, \nonumber \\ 
\beta_3 & = & e^{\mathrm{i}\phi_3} - e^{\mathrm{i}\theta_1} - e^{\mathrm{i}\theta_4} + 1,~~ \gamma_1 = e^{\mathrm{i}\varphi_1} - e^{\mathrm{i}\phi_1} -e^{\mathrm{i}\phi_2} + e^{\mathrm{i}\theta_1}, \label{uc43}  
\end{eqnarray}
for arbitrary angles $\theta_1$, $\theta_4$, $\phi_1$, $\phi_2$ and $\varphi_1$. The eigenvalues at these unitary solutions are given by $\{1_{(4)}, \pm e^{\frac{\mathrm{i}}{2}(\theta_1 + \theta_4)}_{(4)}, e^{\mathrm{i}\phi_3}_{(4)}\}$. 

 As in the $(2,4,2)$ case, there are 64 real unitary generalized $R$-matrices and we write down one unitary solution from each set of eigenvalue.
When the parameters are complex we obtain a family of unitary solutions that generate the full braid group.

\paragraph{Eigenvalues $\in \{-1_{(4)},1_{(12)}\}$}
When $\left(\alpha_1, \alpha_4, \beta_1, \beta_2, \beta_3, \gamma_1\right) = \left(-2,-2,0,0,4,2\right)$, the matrix reads
\begin{equation}
\frac{1}{4}\left(\tiny{\begin{array}{cccccccc|cccccccc} 1 & 0 & 1 & 0 & 1 & 0 & 1 & 0 & 0 & 3 & 0 & -1 & 0 & -1 & 0 & -1 \\ 0 & 3 & 0 & -1 & 0 & -1 & 0 & -1 & 1 & 0 & 1 & 0 & 1 & 0 & 1 & 0 \\ 1 & 0 & 1 & 0 & 1 & 0 & 1 & 0 & 0 & -1 & 0 & 3 & 0 & -1 & 0 & -1 \\ 0 & -1 & 0 & 3 & 0 & -1 & 0 & -1 & 1 & 0 & 1 & 0 & 1 & 0 & 1 & 0 \\ 1 & 0 & 1 & 0 & 1 & 0 & 1 & 0 & 0 & -1 & 0 & -1 & 0 & 3 & 0 & -1 \\ 0 & -1 & 0 & -1 & 0 & 3 & 0 & -1 & 1 & 0 & 1 & 0 & 1 & 0 & 1 & 0 \\ 1 & 0 & 1 & 0 & 1 & 0 & 1 & 0 & 0 & -1 & 0 & -1 & 0 & -1 & 0 & 3 \\ 0 & -1 & 0 & -1 & 0 & -1 & 0 & 3 & 1 & 0 & 1 & 0 & 1 & 0 & 1 & 0 \\ \hline 0 & 1 & 0 & 1 & 0 & 1 & 0 & 1 & 3 & 0 & -1 & 0 & -1 & 0 & -1 & 0 \\ 3 & 0 & -1 & 0 & -1 & 0 & -1 & 0 & 0 & 1 & 0 & 1 & 0 & 1 & 0 & 1 \\ 0 & 1 & 0 & 1 & 0 & 1 & 0 & 1 & -1 & 0 & 3 & 0 & -1 & 0 & -1 & 0 \\ -1 & 0 & 3 & 0 & -1 & 0 & -1 & 0 & 0 & 1 & 0 & 1 & 0 & 1 & 0 & 1 \\ 0 & 1 & 0 & 1 & 0 & 1 & 0 & 1 & -1 & 0 & -1 & 0 & 3 & 0 & -1 & 0 \\ -1 & 0 & -1 & 0 & 3 & 0 & -1 & 0 & 0 & 1 & 0 & 1 & 0 & 1 & 0 & 1 \\ 0 & 1 & 0 & 1 & 0 & 1 & 0 & 1 & -1 & 0 & -1 & 0 & -1 & 0 & 3 & 0 \\ -1 & 0 & -1 & 0 & -1 & 0 & 3 & 0 & 0 & 1 & 0 & 1 & 0 & 1 & 0 & 1 \end{array}}\right).
\end{equation}   
This generates $ \frac{1}{4}\left[\ket{0000} + \ket{0010} + \ket{0100} + \ket{0110} + 3\ket{1001} - \ket{1011} - \ket{1101} - \ket{1111}\right]$ from $\ket{0000}$.

\paragraph{Eigenvalues $\in \{-1_{(8)},1_{(8)}\}$}

The solution for $\left(\alpha_1, \alpha_4, \beta_1, \beta_2, \beta_3, \gamma_1\right) = \left(-2,-2,0,0,2,0\right)$ reads
\begin{equation}
\frac{1}{2}\left(\tiny{\begin{array}{cccccccc|cccccccc} -1 & -1 & 0 & 0 & 0 & 0 & 0 & 0 & -1 & 1 & 0 & 0 & 0 & 0 & 0 & 0 \\ -1 & 1 & 0 & 0 & 0 & 0 & 0 & 0 & -1 & -1 & 0 & 0 & 0 & 0 & 0 & 0 \\ 0 & 0 & -1 & -1 & 0 & 0 & 0 & 0 & 0 & 0 & -1 & 1 & 0 & 0 & 0 & 0 \\ 0 & 0 & -1 & 1 & 0 & 0 & 0 & 0 & 0 & 0 & -1 & -1 & 0 & 0 & 0 & 0 \\ 0 & 0 & 0 & 0 & -1 & -1 & 0 & 0 & 0 & 0 & 0 & 0 & -1 & 1 & 0 & 0 \\ 0 & 0 & 0 & 0 & -1 & 1 & 0 & 0 & 0 & 0 & 0 & 0 & -1 & -1 & 0 & 0 \\ 0 & 0 & 0 & 0 & 0 & 0 & -1 & -1 & 0 & 0 & 0 & 0 & 0 & 0 & -1 & 1 \\ 0 & 0 & 0 & 0 & 0 & 0 & -1 & 1 & 0 & 0 & 0 & 0 & 0 & 0 & -1 & -1 \\ \hline -1 & -1 & 0 & 0 & 0 & 0 & 0 & 0 & 1 & -1 & 0 & 0 & 0 & 0 & 0 & 0 \\ 1 & -1 & 0 & 0 & 0 & 0 & 0 & 0 & -1 & -1 & 0 & 0 & 0 & 0 & 0 & 0 \\ 0 & 0 & -1 & -1 & 0 & 0 & 0 & 0 & 0 & 0 & 1 & -1 & 0 & 0 & 0 & 0 \\ 0 & 0 & 1 & -1 & 0 & 0 & 0 & 0 & 0 & 0 & -1 & -1 & 0 & 0 & 0 & 0 \\ 0 & 0 & 0 & 0 & -1 & -1 & 0 & 0 & 0 & 0 & 0 & 0 & 1 & -1 & 0 & 0 \\ 0 & 0 & 0 & 0 & 1 & -1 & 0 & 0 & 0 & 0 & 0 & 0 & -1 & -1 & 0 & 0 \\ 0 & 0 & 0 & 0 & 0 & 0 & -1 & -1 & 0 & 0 & 0 & 0 & 0 & 0 & 1 & -1 \\ 0 & 0 & 0 & 0 & 0 & 0 & 1 & -1 & 0 & 0 & 0 & 0 & 0 & 0 & -1 & -1 \end{array}}\right).
\end{equation}   
This generates $ \frac{1}{2}\left[-\ket{0000} - \ket{0001} - \ket{1000} + \ket{1001} \right]$
from $\ket{0000}$.

\paragraph{Eigenvalues $\in \{-\mathrm{i}_{(4)}, \mathrm{i}_{(4)},1_{(8)}\}$}

The solution for $\left(\alpha_1, \alpha_4, \beta_1, \beta_2, \beta_3, \gamma_1\right) = \left(-2,0,0,0,2,2\right)$ reads
\begin{equation}
\frac{1}{4}\left(\tiny{\begin{array}{cccccccc|cccccccc} 2 & 1 & 0 & -1 & 0 & -1 & 0 & -1 & -1 & 2 & 1 & 0 & 1 & 0 & 1 & 0 \\ -1 & 2 & 1 & 0 & 1 & 0 & 1 & 0 & 2 & 1 & 0 & -1 & 0 & -1 & 0 & -1 \\ 0 & -1 & 2 & 1 & 0 & -1 & 0 & -1 & 1 & 0 & -1 & 2 & 1 & 0 & 1 & 0 \\ 1 & 0 & -1 & 2 & 1 & 0 & 1 & 0 & 0 & -1 & 2 & 1 & 0 & -1 & 0 & -1 \\ 0 & -1 & 0 & -1 & 2 & 1 & 0 & -1 & 1 & 0 & 1 & 0 & -1 & 2 & 1 & 0 \\ 1 & 0 & 1 & 0 & -1 & 2 & 1 & 0 & 0 & -1 & 0 & -1 & 2 & 1 & 0 & -1 \\ 0 & -1 & 0 & -1 & 0 & -1 & 2 & 1 & 1 & 0 & 1 & 0 & 1 & 0 & -1 & 2 \\ 1 & 0 & 1 & 0 & 1 & 0 & -1 & 2 & 0 & -1 & 0 & -1 & 0 & -1 & 2 & 1 \\ \hline 1 & 2 & -1 & 0 & -1 & 0 & -1 & 0 & 2 & -1 & 0 & 1 & 0 & 1 & 0 & 1 \\ 2 & -1 & 0 & 1 & 0 & 1 & 0 & 1 & 1 & 2 & -1 & 0 & -1 & 0 & -1 & 0 \\ -1 & 0 & 1 & 2 & -1 & 0 & -1 & 0 & 0 & 1 & 2 & -1 & 0 & 1 & 0 & 1 \\ 0 & 1 & 2 & -1 & 0 & 1 & 0 & 1 & -1 & 0 & 1 & 2 & -1 & 0 & -1 & 0 \\ -1 & 0 & -1 & 0 & 1 & 2 & -1 & 0 & 0 & 1 & 0 & 1 & 2 & -1 & 0 & 1 \\ 0 & 1 & 0 & 1 & 2 & -1 & 0 & 1 & -1 & 0 & -1 & 0 & 1 & 2 & -1 & 0 \\ -1 & 0 & -1 & 0 & -1 & 0 & 1 & 2 & 0 & 1 & 0 & 1 & 0 & 1 & 2 & -1 \\ 0 & 1 & 0 & 1 & 0 & 1 & 2 & -1 & -1 & 0 & -1 & 0 & -1 & 0 & 1 & 2  \end{array}}\right).
\end{equation}
This generates $$ \frac{1}{4}\left[2\ket{0000} - \ket{0001} + \ket{0011} + \ket{0101} + \ket{0111} + \ket{1000} + 2\ket{1001} - \ket{1010} - \ket{1100} - \ket{1110}\right]$$
from $\ket{0000}$.

\paragraph{Eigenvalues $\in \{-\mathrm{i}_{(4)}, \mathrm{i}_{(4)},-1_{(4)}, 1_{(4)}\}$}

The solution for $\left(\alpha_1, \alpha_4, \beta_1, \beta_2, \beta_3, \gamma_1\right) = \left(-2,0,0,2,0,0\right)$ reads
\begin{equation}
\frac{1}{2}\left(\tiny{\begin{array}{cccccccc|cccccccc} 0 & -1 & 0 & -1 & 0 & 0 & 0 & 0 & -1 & 0 & 1 & 0 & 0 & 0 & 0 & 0 \\ -1 & 0 & 1 & 0 & 0 & 0 & 0 & 0 & -1 & 0 & -1 & 0 & 0 & 0 & 0 & 0 \\ 0 & -1 & 0 & -1 & 0 & 0 & 0 & 0 & 1 & 0 & -1 & 0 & 0 & 0 & 0 & 0 \\ 1 & 0 & -1 & 0 & 0 & 0 & 0 & 0 & 0 & -1 & 0 & -1 & 0 & 0 & 0 & 0 \\ 0 & 0 & 0 & 0 & 0 & -1 & 0 & -1 & 0 & 0 & 0 & 0 & -1 & 0 & 1 & 0 \\ 0 & 0 & 0 & 0 & -1 & 0 & 1 & 0 & 0 & 0 & 0 & 0 & 0 & -1 & 0 & -1 \\ 0 & 0 & 0 & 0 & 0 & -1 & 0 & -1 & 0 & 0 & 0 & 0 & 1 & 0 & -1 & 0 \\ 0 & 0 & 0 & 0 & 1 & 0 & -1 & 0 & 0 & 0 & 0 & 0 & 0 & -1 & 0 & -1 \\ \hline -1 & 0 & -1 & 0 & 0 & 0 & 0 & 0 & 0 & -1 & 0 & 1 & 0 & 0 & 0 & 0 \\ 0 & -1 & 0 & 1 & 0 & 0 & 0 & 0 & -1 & 0 & -1 & 0 & 0 & 0 & 0 & 0 \\ -1 & 0 & -1 & 0 & 0 & 0 & 0 & 0 & 0 & 1 & 0 & -1 & 0 & 0 & 0 & 0 \\ 0 & 1 & 0 & -1 & 0 & 0 & 0 & 0 & -1 & 0 & -1 & 0 & 0 & 0 & 0 & 0 \\ 0 & 0 & 0 & 0 & -1 & 0 & -1 & 0 & 0 & 0 & 0 & 0 & 0 & -1 & 0 & 1 \\ 0 & 0 & 0 & 0 & 0 & -1 & 0 & 1 & 0 & 0 & 0 & 0 & -1 & 0 & -1 & 0 \\ 0 & 0 & 0 & 0 & -1 & 0 & -1 & 0 & 0 & 0 & 0 & 0 & 0 & 1 & 0 & -1 \\ 0 & 0 & 0 & 0 & 0 & 1 & 0 & -1 & 0 & 0 & 0 & 0 & -1 & 0 & -1 & 0  \end{array}}\right).
\end{equation}   
This generates $ \frac{1}{2}\left[-\ket{0001} + \ket{0011} - \ket{1000} - \ket{1010} \right]$
from $\ket{0000}$.

\subsection{4-qubits from 2-qubits via Temperley-Lieb generators}\label{4quR_TL}

Up to this point we used the qubit representations of \eqref{xrep1} in writing down the $(d,2,1)$-, $(d,3,2)$-, $(d,4,2)$- and $(d,4,3)$-generalized Yang-Baxter operators. As noted in Sec. \ref{pa}, if we instead use the Temperley-Lieb representation we would obtain the generalized Yang-Baxter operators that solve the $(d,4,2)$-, $(d,6,4)$-, $(d,8,4)$-, and $(d,8,6)$-gYBEs. In what follows we discuss the $(d,4,2)$-generalized $R$-matrices in detail. Note that far-commutativity is satisfied for each of these operators, thus leading to braid group representations. 

By applying the realization \eqref{part-TL} and \eqref{chichi} to \eqref{b21} for the 2-qubit case, we obtain the generalized Yang-Baxter operator for 4-qubits at the sites $2i-1, 2i, 2i+1,2i+2$:
\begin{equation}
R_i = s_{2i-1,\,2i+1}s_{2i,\,2i+2}\left(1 + \alpha e_{2i-1} +\beta e_{2i+1} +\gamma e_{2i-1}e_{2i+1}\right).
\label{b4}
\end{equation}
It is clear that this solves $R_iR_{i+1}R_i= R_{i+1}R_iR_{i+1}$, 
but this should be recognized as the $(d,4,2)$- gYBE instead of $(d,4,1)$-gYBE. 
Note that $R_i$ has a nontrivial support on the sites $2i-1$ to $2i+2$, whereas $R_{i+1}$ has that on the sites $2i+1$ to $2i+4$. 
A shift of the index $i$ of the $R$-matrix by one corresponds to a shift of the sites by two. 
The far-commutativity relation is satisfied in the sense of $R_iR_j=R_jR_i$ for $|i-j|>1$. 

If the Temperley-Lieb generators are expressed by hermitian matrices, we can see that the solution is unitary for 
\begin{equation}
\alpha=\frac{1}{\Delta}(e^{\mathrm{i}\theta}-1),\qquad \beta=\frac{1}{\Delta}(e^{\mathrm{i}\varphi}-1),\qquad
\gamma =\frac{1}{\Delta^2}(e^{\mathrm{i}\phi}-e^{\mathrm{i}\theta}-e^{\mathrm{i}\varphi}+1)
\label{unitary_sol_TL}
\end{equation}
with $\Delta\equiv Q+Q^{-1}$, and $\theta$, $\varphi$ and $\phi$ angles taking any value. 
For real $\alpha$, $\beta$ and $\gamma$, there are 8 unitary points: 
\begin{eqnarray}
(\alpha,\beta,\gamma) & = & \large\{ (0,0,0), \left(0,0,-\frac{2}{\Delta^2}\right), \left(0,-\frac{2}{\Delta},0\right), \left(-\frac{2}{\Delta},0,0\right), \nonumber \\
 & &\left(0,-\frac{2}{\Delta},\frac{2}{\Delta^2}\right), 
 \left(-\frac{2}{\Delta},0,\frac{2}{\Delta^2}\right), \left(-\frac{2}{\Delta},-\frac{2}{\Delta},\frac{2}{\Delta^2}\right), \left(-\frac{2}{\Delta},-\frac{2}{\Delta},\frac{4}{\Delta^2}\right)\large\}. 
 \label{b4_unitary}
\end{eqnarray}

As a representation of the Temperley-Lieb generators for qubits at the sites $i$ and $i+1$, we use the following matrix~\cite{BMNR}
\begin{equation}
e_i=\left(Q\ket{01}-\ket{10}\right)\left(\bra{01}-Q^{-1}\bra{10}\right)=\begin{pmatrix}
0 & 0 & 0 & 0 \\ 0 & Q & -1 & 0 \\ 0 & -1 & Q^{-1} & 0 \\ 0 & 0 & 0 & 0 \end{pmatrix}.
\end{equation}
Then, the eigenvalues of the $R$-matrix (\ref{b4}) with (\ref{unitary_sol_TL}) are $\{1_{(6)},\,-1,\,e^{\mathrm{i}\phi},\,\pm e^{\frac{\mathrm{i}}{2}(\theta +\varphi)}_{(3)}\}$, showing that the $R$-matrix generates the infinite-dimensional braid group. 
Note that these are independent of the parameter $Q$. 
In what follows, we  
discuss the corresponding entangled states for the non-trivial cases of (\ref{b4_unitary}). We obtain the $R$-matrices which are not equivalent with those in the $(d,4,2)$-cases in the previous subsection. 

\paragraph{Case I: $\left(0,0,-\frac{2}{\Delta^2}\right)$}
The solution satisfies $R_i^2=1$, and generates entangled states as 
\begin{eqnarray}
R_i\ket{0101} & = & \left(1-\frac{2Q^2}{\Delta^2}\right)\ket{0101}-\frac{2}{\Delta^2}\ket{1010} +\frac{2Q}{\Delta^2}\left(\ket{0110}+\ket{1001}\right), \nonumber \\
R_i\ket{1010} & = & \left(1-\frac{2Q^{-2}}{\Delta^2}\right)\ket{1010}-\frac{2}{\Delta^2}\ket{0101} +\frac{2Q^{-1}}{\Delta^2}\left(\ket{0110}+\ket{1001}\right), \nonumber \\
R_i\ket{0110} & = & \frac{Q^2+Q^{-2}}{\Delta^2}\ket{1001} -\frac{2}{\Delta^2}\ket{0110} +\frac{2Q}{\Delta^2}\ket{0101} +\frac{2Q^{-1}}{\Delta^2}\ket{1010}, \nonumber \\
R_i\ket{1001} & = & \frac{Q^2+Q^{-2}}{\Delta^2}\ket{0110} -\frac{2}{\Delta^2}\ket{1001} +\frac{2Q}{\Delta^2}\ket{0101} +\frac{2Q^{-1}}{\Delta^2}\ket{1010}.
\label{Case1_b4}
\end{eqnarray}
For the other states, $R_i$ does not generate entanglement. 
We can see that each of the four states in \eqref{Case1_b4} is SLOCC equivalent to 
\begin{equation}
\ket{0000} + \ket{1111} +\lambda\left(\ket{0011} + \ket{1100}\right),
\label{Gabcd}
\end{equation}
with some coefficient $\lambda$. This falls into what is called $G_{abcd}$ in \cite{VDMV} with $a=1+\lambda, b=c=0, d=1-\lambda$, or into what is called the class of ${\rm span}\{0_k\Psi, 0_k\Psi\}$ in \cite{LLSS2}.   For example, the first state in \eqref{Case1_b4} is mapped to \eqref{Gabcd} by successively operating the following two ILOs:
\begin{equation}
1\otimes X \otimes 1\otimes X, \quad {\rm diag}\left(\left(1-\frac{2Q}{\Delta^2}\right)^{-1/4},\left(-\frac{2}{\Delta^2}\right)^{-1/4}\right)^{\otimes 4}. 
\end{equation}
The matrix $R_i$ has eigenvalues $1_{(9)}$ and $-1_{(7)}$, which is inequivalent to any of (\ref{d42_1})-(\ref{d42_4}). 

\paragraph{Case II: $\left(0,-\frac{2}{\Delta},0\right)$}
The solution satisfies $R_i^4=1$ (but $R_i^2, R_i^3\neq 1$), 
and generates entangled states as the form of $(\mbox{the Bell state})\otimes (\mbox{separable 2-qubit state})$. 
The eigenvalues of $R_i$ are $\{-\mathrm{i}_{(3)}, \mathrm{i}_{(3)},-1_{(4)}, 1_{(6)}\}$, which is inequivalent to any of (\ref{d42_1})-(\ref{d42_4}). 

\paragraph{Case III: $\left(-\frac{2}{\Delta},0,0\right)$}
This case provides essentially the same entanglement as the case II, since the two $R$-matrices are connected by swapping sites as 
\begin{equation}
\left.R_i\right|_{{\rm case \,III}}=s_{2i-1,2i+1}s_{2i,2i+1}\left(\left.R_i\right|_{{\rm case \,II}}\right) s_{2i-1,2i+1}s_{2i,2i+1}.
\label{R_swap}
\end{equation}

\paragraph{Case IV: $\left(0,-\frac{2}{\Delta},\frac{2}{\Delta^2}\right)$}
The solution satisfies $R_i^4=1$ (but $R_i^2, R_i^3\neq 1$), and gives entangled states as 
\begin{eqnarray}
R_i\ket{0101} & = & \left(1-\frac{2Q}{\Delta}+\frac{2Q^2}{\Delta^2}\right)\ket{0101} +\frac{2}{\Delta^2}\ket{1010} -\frac{2Q}{\Delta^2}\ket{0110}+\frac{2Q^{-1}}{\Delta^2}\ket{1001}, \nonumber \\
R_i\ket{1010} & = & \left(1-\frac{2Q^{-1}}{\Delta}+\frac{2Q^{-2}}{\Delta^2}\right)\ket{1010} +\frac{2}{\Delta^2}\ket{0101} +\frac{2Q}{\Delta^2}\ket{0110}-\frac{2Q^{-1}}{\Delta^2}\ket{1001}, \nonumber \\
R_i\ket{0110} & = & \left(1-\frac{2Q^{-1}}{\Delta}+\frac{2}{\Delta^2}\right)\ket{1001} +\frac{2}{\Delta^2}\ket{0110} +\frac{2Q^{-1}}{\Delta^2}\left(\ket{0101}-\ket{1010}\right), \nonumber \\
R_i\ket{1001} & = & \left(1-\frac{2Q}{\Delta}+\frac{2}{\Delta^2}\right)\ket{0110} +\frac{2}{\Delta^2}\ket{1001} +\frac{2Q}{\Delta^2}\left(\ket{1010}-\ket{0101}\right).
\label{Case4_b4}
\end{eqnarray}
Operating $R_i$ on any of the four states $\ket{0001}, \ket{0010}, \ket{1101}, \ket{1110}$ generates entangled states like $(\mbox{the Bell state})\otimes (\mbox{separable 2-qubit state})$. 
For the other states, $R_i$ gives product states. 
The four states in \eqref{Case4_b4} are SLOCC equivalent to \eqref{Gabcd}. For example, successive operations of the three ILOs
\begin{equation}
1\otimes X\otimes 1\otimes X, \quad 1\otimes \begin{pmatrix} iQ^{-1} &  \\  & 1 \end{pmatrix} \otimes \begin{pmatrix} -iQ &  \\  &1 \end{pmatrix}\otimes 1, \quad 
{\rm diag}\left(\left(1-\frac{2Q}{\Delta}+\frac{2Q^2}{\Delta^2}\right)^{-1/4}, \left(\frac{2}{\Delta^2}\right)^{-1/4}\right)^{\otimes 4}
\end{equation}
map the first state in \eqref{Case4_b4} to \eqref{Gabcd}. 
The eigenvalues of $R_i$ are 
$\{-\mathrm{i}_{(3)}, \mathrm{i}_{(3)},-1_{(3)}, 1_{(7)}\}$, which is inequivalent to any of (\ref{d42_1})-(\ref{d42_4}). 

\paragraph{Case V: $\left(-\frac{2}{\Delta},0,\frac{2}{\Delta^2}\right)$}
This case is essentially the same as the case IV, because 
\begin{equation}
\left.R_i\right|_{{\rm case \,V}}=s_{2i-1,2i+1}s_{2i,2i+1}\left(\left.R_i\right|_{{\rm case \,IV}}\right) s_{2i-1,2i+1}s_{2i,2i+1}.
\label{R_swap2}
\end{equation}

\paragraph{Case VI: $\left(-\frac{2}{\Delta},-\frac{2}{\Delta},\frac{2}{\Delta^2}\right)$}
The solution satisfies $R_i^2=1$, and generates entangled states as 
\begin{eqnarray}
R_i\ket{0101} & = & \left(1-\frac{4Q}{\Delta}+\frac{2Q^2}{\Delta^2}\right)\ket{0101} +\frac{2}{\Delta^2}\ket{1010} +\frac{2Q^{-1}}{\Delta^2}\left(\ket{0110} + \ket{1001}\right), \nonumber \\
R_i\ket{1010} & = & \left(1-\frac{4Q^{-1}}{\Delta}+\frac{2Q^{-2}}{\Delta^2}\right)\ket{1010} +\frac{2}{\Delta^2}\ket{0101} +\frac{2Q}{\Delta^2}\left(\ket{0110} + \ket{1001}\right), \nonumber \\
R_i\ket{0110} & = & \left(-1+\frac{2}{\Delta^2}\right)\ket{1001} +\frac{2}{\Delta^2}\ket{0110} +\frac{2Q}{\Delta^2}\ket{1010} + \frac{2Q^{-1}}{\Delta^2}\ket{0101}, \nonumber \\
R_i\ket{1001} & = & \left(-1+\frac{2}{\Delta^2}\right)\ket{0110} +\frac{2}{\Delta^2}\ket{1001} +\frac{2Q}{\Delta^2}\ket{1010} + \frac{2Q^{-1}}{\Delta^2}\ket{0101}, 
\end{eqnarray}
which are SLOCC equivalent to \eqref{Gabcd}. 
The states $\ket{0001}$, $\ket{1110}$ and their permutations with respect to the sites provide the direct product of the Bell state and a separable 2-qubit state by acting with $R_i$. 
For the other states, $R_i$ gives product states.
The $R_i$ has the eigenvalues $1_{(9)}$  and $-1_{(7)}$, which is inequivalent to any of (\ref{d42_1})-(\ref{d42_4}). 

\paragraph{Case VII: $\left(-\frac{2}{\Delta},-\frac{2}{\Delta},\frac{4}{\Delta^2}\right)$}
Since the solution in this case is factorized as 
\begin{equation}
R_i=s_{2i-1,\,2i+1}\left(1-\frac{2}{\Delta}e_{2i-1}\right)s_{2i\,2i+2}\left(1-\frac{2}{\Delta}e_{2i+1}\right),
\end{equation}
it is easy to see that $R_i^2=1$, and the entangled states obtained fall into a class of the direct product of the two Bell states or the direct product of the Bell state and a separable 2-qubit state.  
The eigenvalues of $R_i$ are $1_{(10)}$  and $-1_{(6)}$, which is inequivalent to any of (\ref{d42_1})-({\ref{d42_4}). 

Although various inequivalent solutions would be obtained by choosing other representations of the Temperley-Lieb algebra, we leave this issue as a future subject.

\subsection{Algorithm for multi-qubit generalized $R$-matrices}\label{mquR}

The ansatze in \eqref{b21}, \eqref{nb31}, \eqref{nb41}, and \eqref{nb42} act as a guide for constructing the generalized $R$-matrices in the multi-qubit case, while using the available operators: $s_j~j=i,\cdots, i+m-2$ and $p_j~j=i,\cdots, i+m-1$. Throughout we assume that $2l\geq m$ in order to ensure far-commutativity  in addition to obeying the $(d,m,l)$-gYBE. 

The generalized $R$-matrix that satisfies the $(d,m,l)$-gYBE and is made up of products of the $p_i$ operators is 
\begin{eqnarray}
R_i & = & s_{i, i+l}\Big[1 + \sum_{r=1}^{m-1}\Big(\sum_{\substack{k_1,\cdots, k_{r-1}=1\\ 0<k_1<\cdots< k_{r-1}<l}}^{m-1}\alpha^{(r)}_{0, k_1, \cdots, k_{r-1}}p_i\prod_{j=1}^{r-1}p_{i+k_j}   + \sum_{\substack{k_1,\cdots, k_{r-1}=1\\ 0<k_1<\cdots< k_{r-1}<l}}^{m-1} \alpha^{(r)}_{l, k_1, \cdots, k_{r-1}}p_{i+l}\prod_{j=1}^{r-1} p_{i+k_j}  \nonumber \\
&  & \hskip 3cm + \sum_{\substack{k_1,\cdots, k_{r-2}=1\\ 0<k_1<\cdots< k_{r-2}<l}}^{m-1}\alpha^{(r)}_{0, k_1, \cdots, k_{r-2}, l} p_ip_{i+l}\prod_{j=1}^{r-2} p_{i+k_j}\Big)  +  \alpha^{(m)}_{0,1,\cdots, m-1}\prod_{j=0}^{m-1}p_{i+j} \Big],\cr
&&
\end{eqnarray}
where $s_{i, i+l} = s_{i+l-1}\cdots s_{i+1}s_is_{i+1}\cdots s_{i+l-1}$ swaps the sites $i$ and $i+l$. 
For $r=1$, $\prod_{j=1}^{r-1}p_{i+k_j}$ and $\prod_{j=1}^{r-2}p_{i+k_j}$ should be regarded as $1$ and $0$, respectively. 
 
The coefficients can be determined by requiring this to satisfy $R_iR_{i+l}R_i=R_{i+l}R_iR_{i+l}$. This computation can be done analytically, but it is tedious. At the moment we do not have a general analytic expression for the generalized $R$-matrix, but this algorithm works as studied in the cases of three and four qubits. As seen in those cases, we expect to obtain a family of both unitary and non-unitary generalized $R$-matrices. The image of the braid group representations using the non-unitary matrices and the complex unitary matrices will be infinite, whereas the image of the braid group representations of the real unitary matrices is expected to be finite. 

Analogous generalized $R$-matrices solving the $(d,m,l)$-gYBE can be constructed using the operators $s_{i, i+l}$ and the powers of $p_{k_1, k_2}$ where at least one of $k_1, k_2$ is either $i$ or $i+l$. We omit the general expression for such an operator here as it is straightforward. 

\section{Comparison with known generalized $R$-matrices}\label{compR}

The known unitary 3-qubit generalized $R$-matrices are the GHZ matrix obtained from the extraspecial 2-group generators in \cite{r1}, the generalization of the Rowell solutions in \cite{rchen}, and the solutions obtained from ribbon fusion categories in \cite{wk, w2}. We now show that the four unitary 3-qubit solutions in Table \ref{ntable1a} are inequivalent to all of the solutions above. 

\paragraph{Non-equivalence to the GHZ matrix}

The GHZ matrix that solves the $(2,3,2)$-gYBE is given by 
\begin{equation}
R_\textrm{GHZ} = \frac{1}{\sqrt{2}}~\left(\begin{array}{cccccccc} 1 & 0 & 0 & 0 & 0 & 0 & 0 & 1 \\ 0 & 1 & 0 & 0 & 0 & 0 & 1 & 0 \\ 0 & 0 & 1 & 0 &  & 1 & 0 & 0 \\ 0 & 0 & 0 & 1 & 1 & 0 & 0 & 0 \\  0 & 0 & 0 & -1 & 1 & 0 & 0 & 0 \\ 0 & 0 & -1 & 0 & 0 & 1 & 0 & 0 \\ 0 & -1 & 0 & 0 & 0 & 0 & 1 & 0 \\ -1 & 0 & 0 & 0 & 0 & 0 & 0 & 1 \end{array}\right)
\end{equation}
and has eigenvalues $e^{\pm\mathrm{i}\frac{\pi}{4}}$, both with multiplicity 4. 
The eigenvalues of the complex unitary 3-qubit matrices of the form (\ref{nb31}) have eigenvalues that cannot be mapped to the eigenvalues of the GHZ matrix either by an inversion or by a scalar multiplication and thus we conclude that at the unitary points these solutions are inequivalent to the GHZ matrix. 

At the same time the eigenvalues of the four real unitary 3-qubit matrices in Table \ref{ntable1a} cannot be mapped to these eigenvalues by a scalar multiplication or an inversion, leading us to the conclusion that those matrices cannot be possibly equivalent to the GHZ matrix. 

\paragraph{Comparison with the generalized Rowell solutions}

In \cite{rchen} the Rowell solutions of the form $\left(\begin{array}{cc}X & 0 \\ 0 & Y\end{array}\right)= X \oplus Y$, are generalized to obtain three families of solutions. They have eigenvalues from the sets $\{e^{-\mathrm{i}\frac{\pi}{12}}, e^{-\mathrm{i}\frac{\pi}{12}}, e^{\mathrm{i}\frac{7\pi}{12}}, e^{\mathrm{i}\frac{7\pi}{12}}\}$, $\{e^{-\mathrm{i}\frac{\pi}{4}}, -e^{-\mathrm{i}\frac{\pi}{4}}, e^{\mathrm{i}\frac{\pi}{4}}, e^{\mathrm{i}\frac{\pi}{4}}\}$ and $\{e^{-\mathrm{i}\frac{\pi}{4}}, e^{-\mathrm{i}\frac{\pi}{4}}, e^{\mathrm{i}\frac{\pi}{4}}, e^{\mathrm{i}\frac{\pi}{4}}\}$, respectively. However, these solve the $(2,3,1)$-gYBEs and hence cannot be compared to the solutions in this paper, which solve instead the $(2,3,2)$-gYBE.

\paragraph{Comparison with the Kitaev-Wang solutions}

The Kitaev-Wang solutions in \cite{wk} and their qubit realizations in \cite{w2} solve the $(d,3,1)$-gYBE, whereas the methods presented in this paper generate the generalized Yang-Baxter operators that solve the $(d,3,2)$-gYBE. Thus we cannot compare the two generalized $R$-matrices. 

\section{Outlook}\label{out}

In this paper we have used the notion of partition algebras to introduce a solution-generating technique to obtain parameter-independent generalized $R$-matrices. This is quite remarkable, since solving the gYBE is a notoriously difficult task. This is especially true for the parameter-independent gYBE, for which very few solutions are known in the literature. In some recent work \cite{pdd}, we have focused on the parameter-dependent case, using supersymmetry algebras instead of partition algebras. In that case, the relation between $R$-matrices and braid group representations was not very clear, so that the present work should be considered as an improvement of our previous analysis. 

Improved as it may be, we need however to remark that the method based on partition algebras has certain limitations. The main issue is that not all SLOCC classes of entangled states seem to be obtainable from the generalized $R$-matrices via partition algebras. In particular, we do not obtain the W-state SLOCC class of a 3-qubit system in the case that $R$ is a unitary braid operator. We suspect that this absence of the W-state class is not peculiar to 3-qubit systems, and it might extend to the multi-qubit case as well. It would be interesting to check whether this is true, although it would represent a very laborious computation. 

Understanding such a distinction of the W states, in particular from the GHZ states, would be relevant to various applications in quantum information and quantum computing. The GHZ states become unentangled bipartite mixed states after tracing out one qubit, while the W states are still entangled. Thus, the GHZ states and their multipartite  analogs are fragile against particle loss and suitable for quantum secret sharing. On the other hand, the W states  and their multiqubit versions are robust against loss with possible application to quantum memories, multiparty  quantum network protocols and universal quantum cloning machines.   

A new technique to construct solutions to generalized Yang-Baxter equation is presented via interplay with $k$-graphs in \cite{Dyang}. It will also interesting to do similar analysis on the entanglement generated through possible novel solutions obtained from the technique.  

Of course, the physical realization of braiding operators is of the utmost interest. A natural next step would then be to try to identify the anyons corresponding to these representations and study their computational power. This could possibly help identifying the unitary modular tensor categories that describe these anyons \cite{umtc}.

On a complementary direction, one could construct new integrable spin chains upon Baxterizing the 2-qubit $R$-matrices. In particular, using the Temperley-Lieb representations of the 2-qubit $R$-matrices one could obtain new 4-site interaction spin chains that are integrable. This is something on which we hope to report at some point in the future.


\paragraph{Note added} 
After the conclusion of this work, we have found nonunitary braid operators corresponding to W states \cite{Padmanabhan:2020frr}, which were left out of the present analysis.  We have managed this by using partition algebras for W states in a 3-qubit space and extraspecial 2-groups for the 4-qubit space case.


\subsection*{Acknowledgements}
We thank Z. Wang for his comments on the manuscript. PP and FS are supported by the Institute for Basic Science in Korea (IBS-R024-Y1, IBS-R018-D1). DT is supported in part by the INFN grant {\it Gauge and String Theory (GAST)}.



\begin{thebibliography}{99}
\addtolength{\parskip}{-1ex}
{\small 

\bibitem{pk}
P. K. Aravind, 
{\it Borromean Entanglement of the GHZ state}, 
in R. S. Cohen, M. Horne, J. Stachel (eds.), {\it Potentiality, Entanglement and Passion-at-a-Distance}, Boston Studies in the Philosophy of Science {\bf 194}, Springer (1997),
\href{https://doi.org/10.1007/978-94-017-2732-7}{DOI:10.1007/978-94-017-2732-7}.

\bibitem{sug}
A.~Sugita,
{\it Borromean Entanglement Revisited},
\href{https://arxiv.org/abs/0704.1712}{arXiv:0704.1712 [quant-ph]}.

\bibitem{quinta} 
  G.~M.~Quinta and R.~Andr\'e,
  {\it Classifying quantum entanglement through topological links},
  Phys.\ Rev.\ A {\bf 97}, no. 4, 042307 (2018), 
\href{https://doi.org/10.1103/PhysRevA.97.042307}{DOI:10.1103/PhysRevA.97.042307}
\href{https://arxiv.org/abs/1803.08935}{[arXiv:1803.08935 [quant-ph]]}.
  
\bibitem{lh1}
L. H. Kauffman, S. J. Lomonaco, 
{\it Quantum Entanglement and Topological Entanglement}, 
New J. Phys. 4 73 (2002), 
\href{https://doi.org/10.1088/1367-2630/4/1/373}{DOI:10.1088/1367-2630/4/1/373}
\href{https://arxiv.org/abs/quant-ph/0205137}{[arXiv:quant-ph/0205137]}.

\bibitem{lh2}
L. H. Kauffman, S. J. Lomonaco Jr, 
{\it Braiding Operators are Universal Quantum Gates}, 
New J. Phys. 6 134 (2004),
\href{https://doi.org/10.1088/1367-2630/6/1/134}{DOI:10.1088/1367-2630/6/1/134}
\href{https://arxiv.org/abs/quant-ph/0401090}{[arXiv:quant-ph/0401090]}.

\bibitem{lh3}
Y. Zhang, L. H. Kauffman, M.-L. Ge, 
{\it Universal Quantum Gate, Yang-Baxterization and Hamiltonian}, 
Int. J. of Quantum Inf., Vol. 3, No. 4 (2005) 669-678, 
\href{https://doi.org/10.1142/S0219749905001547}{DOI:10.1142/S0219749905001547}
\href{https://arxiv.org/abs/quant-ph/0412095}{[arXiv:quant-ph/0412095]}.  

\bibitem{lh4}
Y. Zhang, L.H. Kauffman and M.L. Ge, 
{\it Yang-Baxterizations, Universal Quantum Gates and Hamiltonians}, 
Quant. Inf. Proc. 4 (2005) 159-197,
\href{https://doi.org/10.1007/s11128-005-7655-7}{DOI:10.1007/s11128-005-7655-7} 
\href{https://arxiv.org/abs/quant-ph/0502015}{[arXiv: quant-ph/0502015]}.

\bibitem{lh5}
L. H. Kauffman, E. Mehrotra, 
{\it Topological Aspects of Quantum Entanglement},  
Quantum Inf. Process (2019) 18: 76,
\href{https://doi.org/10.1007/s11128-019-2191-z}{DOI:10.1007/s11128-019-2191-z} 
\href{https://arxiv.org/abs/1611.08047}{[arXiv:1611.08047 [math.GT]]}. 

\bibitem{ste}
G. Alagic, M. Jarret, S. P. Jordan, 
{\it Yang-Baxter operators need quantum entanglement to distinguish knots}, 
J. Phys. A, 49 075203 (2016),
\href{https://doi.org/10.1088/1751-8113/49/7/075203}{DOI:10.1088/1751-8113/49/7/075203} 
\href{https://arxiv.org/abs/1507.05979}{[arXiv:1507.05979 [quant-ph]]}. 

\bibitem{r1} 
E. C. Rowell, Y. Zhang, Y. S. Wu, M. L. Ge, 
{\it Extraspecial Two-Groups, Generalized Yang-Baxter Equations and Braiding Quantum Gates}, 
Quant. Inf. Comput.10:685-702, 2010
\href{https://doi.org/10.26421/QIC10.7-8}{DOI:10.26421/QIC10.7-8}
\href{https://arxiv.org/abs/0706.1761}{[arXiv:0706.1761 [quant-ph]]}.

\bibitem{r2} 
J. Franko, E. C. Rowell and Z. Wang, 
{\it Extraspecial 2-Groups and Images of Braid Group Representations}, 
J. Knot Theory Ramifications, 15 (2006) 413-428,
\href{https://doi.org/10.1142/S0218216506004580}{DOI:10.1142/S0218216506004580} 
\href{https://arxiv.org/abs/math/0503435}{[arXiv:math.RT/0503435]}.

\bibitem{wk}
A. Kitaev, Z. Wang, 
{\it Solutions to generalized Yang-Baxter equations via ribbon fusion categories}, 
GTM 18 (2012) 191-197,
\href{https://doi.org/10.2140/gtm.2012.18.191}{DOI:10.2140/gtm.2012.18.191} 
\href{https://arxiv.org/abs/1203.1063}{[arXiv:1203.1063 [math.QA]]}.

\bibitem{w2}
J. F. Vasquez, Z. Wang, H. M. Wong,
{\it Qubit representations of the braid groups from generalized Yang-Baxter matrices},
Quantum Information Processing 15(7) 2016,
\href{https://doi.org/10.1007/s11128-016-1313-0}{DOI:10.1007/s11128-016-1313-0}
\href{https://arxiv.org/abs/1602.08536}{[arXiv:1602.08536 [math.QA]]}.

\bibitem{lh6}
L. H. Kauffman,
{\it Knot Logic and Topological Quantum Computing with Majorana Fermions},
in J. Chubb, J. Chubb, A. Eskandarian, and V. Harizanov (eds.), {\it Lecture Notes in Logic},  Cambridge University Press,
\href{https://doi.org/10.1017/CBO9781139519687.012}{DOI:10.1017/CBO9781139519687.012}
\href{https://arxiv.org/abs/1301.6214}{[arXiv:1301.6214 [quant-ph]]}.

\bibitem{lh7}
L. H. Kauffman, S. J. Lomonaco Jr,
{\it $q$-Deformed Spin Networks, Knot Polynomials and Anyonic Topological Quantum Computation},
J. Knot Theory Ramifications 16(3) 2007,
\href{https://doi.org/10.1142/S0218216507005282}{DOI:10.1142/S0218216507005282}
\href{https://arxiv.org/abs/quant-ph/0606114}{[arXiv:quant-ph/0606114v3]}.

\bibitem{pdd}
P. Padmanabhan, F. Sugino, D. Trancanelli,
{\it Quantum entanglement, supersymmetry, and the generalized Yang-Baxter equation},
Quantum Information and Computation Vol. 20, No. 1\& 2, (2020) 0037-0064,
\href{https://doi.org/10.26421/QIC20.1-2}{DOI:10.26421/QIC20.1-2}
\href{https://arxiv.org/abs/1911.02577}{[arXiv:1911.02577 [quant-ph]]}.

\bibitem{hietarinta2}
J.~Hietarinta,
{\it All solutions to the constant quantum Yang-Baxter equation in two dimensions,} 
Physics Letters A {\bf 165} (1992) 245-251,
\href{https://doi.org/10.1016/0375-9601(92)90044-M}{DOI:10.1016/0375-9601(92)90044-M}.

\bibitem{hietarinta}
J.~Hietarinta, 
{\it The upper triangular solutions to the three-state constant quantum Yang-Baxter equation}, 
J. Physics A: General Physics 26(23):7077 (1999),
\href{https://doi.org/10.1088/0305-4470/26/23/044}{DOI:10.1088/0305-4470/26/23/044}
\href{https://arxiv.org/abs/solv-int/9306001}{[arXiv:solv-int/9306001]}.

\bibitem{dye}
H.~A.~Dye,
{\it Unitary Solutions to the Yang-Baxter Equation in Dimension Four},
 Quantum Information Processing 2, 117-152 (2003),
\href{https://doi.org/10.1023/A:1025843426102}{DOI:10.1023/A:1025843426102}
\href{https://arxiv.org/abs/quant-ph/0211050}{[arXiv:quant-ph/0211050]}.

\bibitem{jfranko}
J. M. Franko,
{\it Braid Group Representations arising from the Yang Baxter Equation},
J.  Knot Theory Ramifications 19(4) 2010,
\href{https://doi.org/10.1142/S021821651000798X}{DOI:10.1142/S021821651000798X}
\href{https://arxiv.org/abs/0807.4138}{[arXiv:0807.4138 [math.GT]]}.

 \bibitem{rchen} 
 R. Chen, 
{\it Generalized Yang-Baxter Equations and Braiding Quantum Gates}, 
J.  Knot Theory Ramifications 21(9) 2011,
\href{https://doi.org/10.1142/S0218216512500873}{DOI:10.1142/S0218216512500873} 
\href{https://arxiv.org/abs/1108.5215}{[arXiv:1108.5215 [math.QA]]}. 

\bibitem{GKRZ}
P.~Gustafson, A.~Kimball, E.~C.~Rowell, Q.~Zhang, 
{\it Braid group representations from twisted tensor products of algebras}, 
Peking Math. Journal,
\href{https://doi.org/10.1007/s42543-020-00023-5}{DOI:10.1007/s42543-020-00023-5}
\href{https://arxiv.org/abs/1906.08153}{[arXiv:1906.08153 [math.QA]]}.  

\bibitem{Jo}
V. F. R. Jones, {\it The Potts model and the symmetry group,} in {\it Subfactors: Proceedings of the Taniguchi Symposium on Operator Algebras, Kyuzeso 1993}, World Scientific Publishing (1994) 259-267.

\bibitem{Ma1}
P. Martin, {\it Potts models and related problems in statistical mechanics,} in 
{\it Series on Advances in Statistical Mechanics}, World Scientific Publishing (1991),
\href{https://doi.org/10.1142/0983}{DOI:10.1142/0983}.

\bibitem{pm}
P. Martin,
{\it Temperley-Lieb Algebras for Non-Planar Statistical Mechanics - The Partition Algebra Construction},
J. Knot Theory Ramifications, 3 (1994) 51-82,
\href{https://doi.org/10.1142/S0218216594000071}{DOI:10.1142/S0218216594000071}.

\bibitem{Ma3} 
P. Martin, {\it The structure of the partition algebras,} J. Algebra {\bf 183} (1996) 319-358,
\href{https://doi.org/10.1006/jabr.1996.0223}{DOI:10.1006/jabr.1996.022}.

\bibitem{Ma4}
P. Martin, {\it The partition algebra and the Potts model transfer matrix spectrum in high dimensions,} J. Phys. A: Math. Gen. {\bf 33} (2000) 3669-3695,
\href{https://doi.org/10.1088/0305-4470/33/19/304}{DOI:10.1088/0305-4470/33/19/304}.

\bibitem{MR} 
P. Martin, G. Rollet, {\it The Potts model representation and a Robinson-Schensted correspondence for the partition algebra,} Compositio Math. {\bf 112} (1998) 237-254,
\href{https://doi.org/10.1023/A:100040041473}{DOI:10.1023/A:100040041473}. 

\bibitem{par}
T. Halverson, A. Ram,
{\it Partition Algebras},
European J. of Combinatorics {\bf 26}, Issue 6 (2005) 869-921,
\href{https://doi.org/10.1016/j.ejc.2004.06.005}{DOI:10.1016/j.ejc.2004.06.005}
\href{https://arxiv.org/abs/math/0401314}{[arXiv:math/0401314 [math.RT]]}.

\bibitem{bry} 
J. L. Brylinski and R. Brylinski, 
{\it Universal Quantum Gates}, in {\it Mathematics of Quantum Computation}, 
Chapman \& Hall/CRC Press (2002),
\href{https://doi.org/10.1201/9781420035377.pt2}{DOI:10.1201/9781420035377.pt2}.

\bibitem{dur} 
W. Dur, G. Vidal, J. I. Cirac, 
{\it Three qubits can be entangled in two inequivalent ways}, 
Phys. Rev. A 62, 062314 (2000),
\href{https://doi.org/10.1103/PhysRevA.62.062314}{DOI:10.1103/PhysRevA.62.062314} 
\href{https://arxiv.org/abs/quant-ph/0005115}{[arXiv:quant-ph/0005115]}.

\bibitem{VDMV}
F. Verstraete, J. Dehaene, B. De Moor, H. Verschelde,
{\it Four qubits can be entangled in nine different ways}, 
Phys. Rev. A 65, 052112 (2002),
\href{https://doi.org/10.1103/PhysRevA.65.052112}{DOI:10.1103/PhysRevA.65.052112} 
\href{https://arxiv.org/abs/quant-ph/0109033}{[arXiv:quant-ph/0109033]}. 

\bibitem{LLSS1}
L. Lamata, J. Leon, D. Salgado, E. Solano,
{\it Inductive classification of multipartite entanglement under SLOCC}, 
Phys. Rev. A 74, 052336 (2006),
\href{https://doi.org/10.1103/PhysRevA.74.052336}{DOI:10.1103/PhysRevA.74.052336} 
\href{https://arxiv.org/abs/quant-ph/0603243}{[arXiv:quant-ph/0603243]}.

\bibitem{LLSS2}
L. Lamata, J. Leon, D. Salgado, E. Solano,
{\it Inductive Entanglement Classification of Four Qubits under SLOCC}, 
Phys. Rev. A 75, 022318 (2007),
\href{https://doi.org/10.1103/PhysRevA.74.052336}{DOI:10.1103/PhysRevA.74.052336} 
\href{https://arxiv.org/abs/quant-ph/0610233}{[arXiv:quant-ph/0610233]}.

\bibitem{LLHL} 
D. Li, X. Li, H. Huang, X. Li,
{\it SLOCC classification for nine families of four-qubits},
Quantum Information and Computation, Vol. 9, No. 9 \& 10 (2009) 0778-0800
\href{https://doi.org/10.26421/QIC9.9-10}{DOI:10.26421/QIC9.9-10}
\href{https://arxiv.org/abs/0712.1876}{[arXiv:0712.1876 [quant-ph]]}.

\bibitem{BMNR} 
M. T. Batchelor, L. Mezincescu, R. I. Nepomechie, V. Rittenberg,
{\it q-deformations of the O(3) symmetric spin-1 Heisenberg chain},
J. Phys. A 23, L141 (1990),
\href{https://doi.org/10.1088/0305-4470/23/4/003}{DOI:10.1088/0305-4470/23/4/003}.

\bibitem{Dyang} 
D. Yang, {\it The interplay between k-graphs and the Yang-Baxter equation,} Journal of Algebra {\bf 451} (2016) 494-525,
\href{https://doi.org/10.1016/j.jalgebra.2016.01.001}{DOI:10.1016/j.jalgebra.2016.01.001}
\href{https://arxiv.org/abs/1506.03117}{[arXiv:1506.03117 [math.QA]]}. 

\bibitem{umtc}
E. Rowell, R. Stong, Z. Wang,
{\it On classification of modular tensor categories},
Comm. Math. Phys. 292 (2009) no. 2, 343-389,
\href{https://doi.org/10.1007/s00220-009-0908-z}{DOI:10.1007/s00220-009-0908-z}
\href{https://arxiv.org/abs/0712.1377}{[arXiv:0712.1377 [math.QA]]}.

\bibitem{Padmanabhan:2020frr}
P.~Padmanabhan, F.~Sugino and D.~Trancanelli,
{\it Generating W states with braiding operators,}
\href{https://arxiv.org/abs/2003.00244}{arXiv:2007.05660 [quant-ph]}.

}
\end{thebibliography}
\end{document}